\documentclass{sig-alternate}

\usepackage{caption}
\usepackage{balance}
\usepackage{verbatim}
\usepackage{multirow}
\usepackage{verbatim}
\usepackage{sparklines}
\usepackage{url}
\usepackage{pifont}
\usepackage{pslatex}

\begin{document}

\title{Chaff from the Wheat : Characterization and Modeling of Deleted Questions on Stack Overflow}

\numberofauthors{2} 
\author{
\alignauthor
Denzil Correa, Ashish Sureka \\ 
\affaddr{Indraprastha Institute of Information Technology IIIT-Delhi }\\ 
\email{\{denzilc, ashish\} @iiitd.ac.in} \\ 
}

%
%
%
%
%

\maketitle
\begin{abstract}
Community based Question Answering (CQA) websites provide dynamic, dedicated platforms to Internet users to seek information via crowdsourced contributions. Stack Overflow is the most popular CQA for programmers on the web with 2.05M users, 5.1M questions and 9.4M answers. Stack Overflow has explicit, detailed guidelines on \emph{how to post} questions and an ebullient moderation community. Despite these precise communications and safeguards, questions posted on Stack Overflow can be extremely off topic or very poor in quality. Such questions can be \emph{deleted} from Stack Overflow at the discretion of experienced community members and moderators. We present the first study of \emph{deleted questions} on Stack Overflow. We divide our study into two parts -- (i) Characterization of \emph{deleted questions} over $\approx$ 5 years (2008-2013) of data,  (ii) Prediction of \emph{deletion} at the time of question creation.


Our characterization study reveals multiple insights on question deletion phenomena. We observe a significant increase in the number of \emph{deleted questions} over time. We find that it takes substantial time to vote a question to be deleted but once voted, the community takes swift action. We also see that question authors delete their questions to salvage reputation points. We notice some instances of accidental deletion of good quality questions but such questions are voted back to be undeleted quickly. We discover a pyramidal structure of question quality on Stack Overflow and find that \emph{deleted questions} lie at the bottom (lowest quality) of the pyramid. We also build a predictive model to detect the \emph{deletion} of question at the creation time. We experiment with 47 features  -- based on \emph{User Profile}, \emph{Community Generated}, \emph{Question Content} and \emph{Syntactic} style -- and report an accuracy of 66\%. Our feature analysis reveals that all four categories of features are important for the prediction task. Our findings reveal important suggestions for content quality maintenance on community based question answering websites. To the best of our knowledge, this is the first large scale study on poor quality (\emph{deleted}) questions on Stack Overflow.

\end{abstract}




\section{Introduction}

\subsection{Research Motivation and Aim}

There has been an increase in the presence of Community Based Question Answering (CQA) website services like Yahoo! Answers, Ask.com and Quora on the Internet. Stack Exchange is a growing network of thematic question-answering websites with each website dedicated to a specific field of expertise~\cite{Jeff-Atwood:2009fk}. It consists of 107 CQA websites with 4.1M users, 7.3M questions, 13.2 M answers and 8.5M visits per day.\footnote{\scriptsize{\url{https://stackexchange.com/about}}} CQA websites on Stack Exchange span across different orthogonal themes like \emph{Technology} (Web Applications, Game Development), \emph{Culture} (Travel, Christianity), \emph{Arts} (Photograph, Scientific Fiction) and \emph{Sciences} (Mathematics, Physics). Stack Overflow is a programming based CQA and the most popular Stack Exchange website consisting of 5.1M questions, 9.4M answers and 2.05 registered users on its website.\footnote{\scriptsize{\url{http://stackoverflow.com/}}}

Stack Overflow has detailed, explicit guidelines on posting questions and it maintains a firm emphasis on following a question-answer format. The community strongly discourages questions which could generate chit-chat, opinions, polls etc. and employs elected moderators to ensure content quality maintenance. Stack Overflow is a free, open (no registration required) website to all users on the Internet and hence, it is a necessity to maintain quality of content on the website ~\cite{agichtein2008finding}. Stack Overflow is driven by the goal to be an exhaustive knowledge base on programming related topics and hence, the community would like to ensure minimal possible \emph{noise} on the website. However, despite of the presence of question posting guidelines and an ebullient moderation community, a significant percentage of questions on Stack Overflow are extremely poor in nature. Questions are a fundamental aspect of any CQA website and the presence of poor quality content may affect user experience. Prior work also shows that poor quality content on a CQA website may drive users away from the platform and adversely affect the traffic~\cite{mamykina2011design}. Therefore, it is important to study poor quality questions and develop mechanisms to minimize them on the website. In this work, we focus our attention on \emph{deleted} questions (poor quality) on Stack Overflow. Questions which are very poor in quality or extremely off topic in nature are deleted from the website. Table~\ref{tab:del-ex} shows examples of deleted questions on Stack Overflow. We see that most of these questions are very poor in quality and of little worth to the community. 

\begin{table}[ht!] \scriptsize
\captionsetup{font=scriptsize, labelfont=bf, textfont=bf}
\centering
\caption{shows examples of deleted 	questions on Stack Overflow}
\begin{tabular}{l|l|l} \hline
\textbf{Id} & \textbf{Title} & \textbf{Score}  \\ \hline
1644242 & Get drive information (free space, etc.) for  & -50 \\
~ & drives on Windows and populate a memo box &  \\
2022741 & plzz can u help me & -9  \\
14771464 & NDTV iPad app screen design & -8 \\
2351964 & I need to see this q fast ..pleash & -8 \\
3077356 & PostgreSQL stupidity & -8 \\
2984348 & hi question about mathematics & -3  \\
\hline
\end{tabular}
\label{tab:del-ex}
\end{table}

Our research aim is to understand the phenomena of \emph{deleted} questions on Stack Overflow to gain insights about the nature of poor quality questions. In addition, we also develop a predictive framework to detect the probability of a question to be deleted at the time of question creation. Such a predictive framework would help the moderators of the Stack Overflow community to detect a poor quality question on Stack Overflow. However, a deleted question on Stack Overflow is an explicit feedback to the question asker that her question does not conform to the guidelines. A predictive framework can convey immediate feedback to the question asker about her question. This may help her to revise her question and improve it in order to avoid deletion. Therefore, prediction of a deleted question at post creation time has two distinct benefits -- (i) An immediate feedback mechanism to the question asker which would serve as an indicator that her question is against the Stack Overflow Q\&A guidelines and (ii) A complementary mechanism to aid community moderators in their daily moderation tasks to identify and delete these extremely poor quality questions from the website.

\subsection{Research Contributions}

We perform the first large scale study on poor quality or \emph{deleted} questions on Stack Overflow. We make the following research contributions -

\begin{itemize}
	\item We analyze deleted questions on Stack Overflow posted over $\approx$5 years and conduct a characterization study. We perform a longitudinal study of deleted questions, community voting patterns and deletion behavior by question owners. We also discover a question quality structure of questions on Stack Overflow. We make our data on \emph{deleted questions} publicly available for research purposes.
	
	\item We develop a predictive model using a machine learning framework to detect a deleted question at the time of question creation. We experiment with 47 features based on four diverse categories and report 66\% accurate predictions. We also perform analysis of our feature space and report features with best discriminatory powers.
\end{itemize}

\section{Related Work}
\label{sec:rel-work}

Stack Overflow is a collaborative question answering Stack Exchange website. The underlying theme of Stack Overflow is programming-related topics and the target audience are software developers, maintenance professionals and programmers. Apart from existing as a question-answering website, the objective of Stack Overflow is to be a comprehensive knowledge base of programming topics. Therefore, Stack Overflow has attracted increasing attention from different research communities like software engineering, human computer interaction, social computing and data mining~\cite{andersonsteering, barua2012developers, bosu2013building, pal2012exploring, ponzanelli2013seahawk}. Researchers have mined the Stack Overflow knowledge-base to extract interesting and unique insights like API usage obstacles, innovation diffusion via URL link sharing,mobile development issues and programming topic trends~\cite{barua2012developers, gomez2013study, linares2013exploratory, wang2013detecting}. Questions on Stack Overflow have received focused attention from researchers. 

Nasehi \emph{et al.} analyze questions on Stack Overflow to understand the quality of a code example~\cite{nasehi2012makes}. They find nine attributes of good questions like concise code, links to extra resources and inline documentation. Wang and Godfrey analyze iOS and Android developer questions on Stack Overflow to detect API usage obstacles~\cite{wang2013detecting}. They used topic models to find a set API classes on iOS and Android documentation which were difficult for developers to understand. Asaduzzaman \emph{et al.} analyze unanswered questions on Stack Overflow and use a machine learning classifier to predict such questions~\cite{asaduzzaman2013answering}. They observe certain characteristics of unanswered questions which include vagueness, homework questions etc. Allamanis and Sutton perform a topic modeling analysis on Stack Overflow questions to combine topics, types and code~\cite{allamanis2013and}. They find that programming languages are a mixture of concepts and questions on Stack Overflow are concerned with the code example rather than the application domain. In contrast to the aforementioned work, our work specifically focuses on quality of content on Stack Overflow.

Measurement of answer quality on CQA has received significant attention using information retrieval models and techniques. Jeon~\emph{et al.} use non-textual question features using a maximum entropy classifier to predict answer quality on Naver, a Korean CQA website~\cite{Jeon:2006:FPQ:1148170.1148212}. Agichtein \emph{et al.} use user relationships, question metadata and content based data to find high quality content on Yahoo! Answers~\cite{agichtein2008finding}. Shah \emph{et al.} build a classifier to predict answer quality on Yahoo! Answers CQA with the usage of question and answer based textual features~\cite{shah2010evaluating}. Most of the previous work focuses on answer quality on large scale CQA websites. Prior work reveals that answer quality is a direct function of question quality~\cite{agichtein2008finding}. Poor quality questions may adversely affect the user experience on the website and therefore, it is important to study such questions~\cite{mamykina2011design}. To this end, Li \emph{et al.} study characteristics of question quality in Yahoo! Answers and find that a Mutual Reinforcement Label Propagation approach based on question plus answer features yield good results~\cite{Li:2012:APQ:2187980.2188200}. Correa and Sureka analyze and predict `closed' questions on Stack Overflow viz. questions which are irrelevant or unfit to the Stack Overflow format~\cite{Correa:2013vn}. In contrast to these previous studies, our current focus is on analysis and prediction of deleted or poor quality questions on Stack Overflow (a programming related CQA). Deleted questions are extremely poor, off-topic in nature and therefore, cease to exist from the website. The properties of deleted questions are different in both topic and content. To the best of our knowledge, this is the first work which studies poor quality questions on a large-scale CQA website like Stack Overflow. 

\section{Deleted Questions on Stack \\ Overflow}
\label{sec:del-ques-so}

\begin{figure}[ht!]
\captionsetup{font=scriptsize, labelfont=bf, textfont=bf}
\centering
\includegraphics[width=\linewidth]{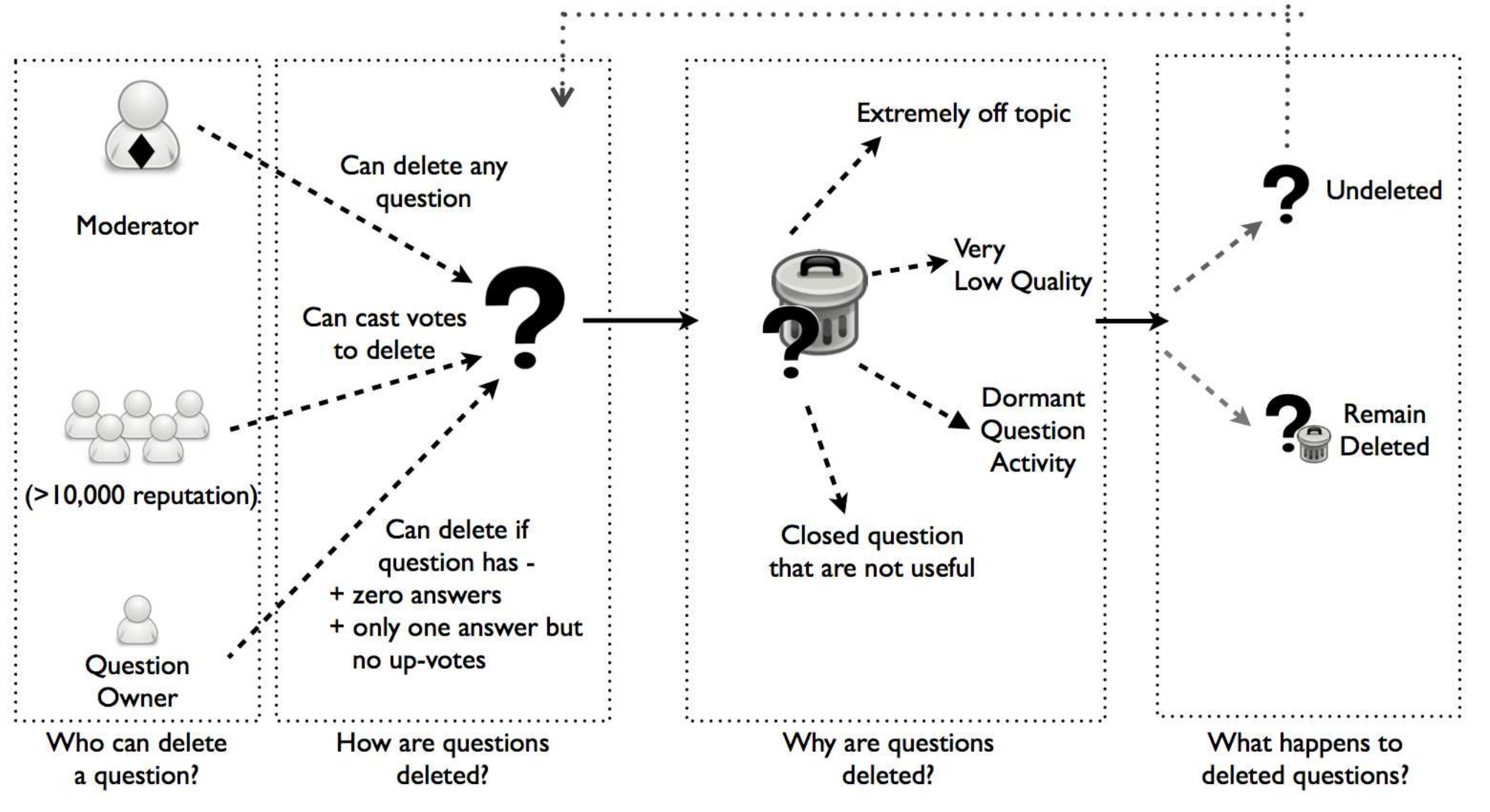}
\caption{shows the details about procedures and community guidelines to delete a question on Stack Overflow.}
\label{fig:del-who-block}
\end{figure}

In this section, we briefly discuss about deleted questions on Stack Overflow. Figure~\ref{fig:del-who-block} summarizes various details about deleted questions on Stack Overflow.

\paragraph{Why are questions deleted?}
The goal of Stack Overflow is to be the most extensive knowledge base of programming related topics. Hence, it would like to keep the \emph{noise} on their website as low as possible. Therefore, questions on Stack Overflow which are extremely off topic or very poor in quality are deleted from the website~\cite{:2012zr}. In addition, questions which are dormant viz. have no activity over a significant period of time are also deleted. Also, `closed' questions (questions which are deemed unfit for Stack Overflow) which do not serve as useful sign points may also be deleted. The \emph{Why} block of Figure~\ref{fig:del-who-block} corresponds to this section. 

\paragraph{Who can delete a question?}
Experienced users with 10,000+ reputation points can cast `delete votes' in order to delete a question. Question authors can delete their own questions if -- (i) the question has zero answers OR (ii) the question has only one answer but no up votes. However, a community elected moderators (super users of the website) can delete any questions at their discretion~\cite{:2008kx}. The \emph{Who} block of Figure~\ref{fig:del-who-block} corresponds to this section. 

\paragraph{How are questions deleted?}

The number of votes required to delete a question scales to the number of answers and the up votes on the answers of those questions. This notwithstanding, a minimum of 3-votes and a maximum of 10-votes are required to delete a question. It must be noted that a deleted question is not physically deleted from the website but \emph{soft deleted}. Moderators, developers and experienced users (10000+ reputation) points are able to view these questions. However, deleted questions do not appear in search results~\cite{:2012zr}. The \emph{How} block of Figure~\ref{fig:del-who-block} corresponds to this section. 

\paragraph{What happens after a question is deleted?}

Once a question is deleted there are two possibilities -- (i) it remains deleted and (ii) it is \emph{undeleted}. The procedure for \emph{undeleting} questions is similar to that of deleting a question. The \emph{What} block of Figure~\ref{fig:del-who-block} corresponds to this section. 

\section{Characterization of Deleted \\ Questions}
\label{sec:char-del-ques}

In this section, we present our findings on deleted questions on Stack Overflow.

\subsection{Dataset Description}

Stack Overflow provides a periodic database dump of all user-generated content under the \emph{Creative Commons Attribute-ShareAlike}~\cite{Atwood:2009uq}. The database dump contains publicly available information of questions, answers, comments, votes and badges from the genesis of Stack Overflow (August 2008) to the release time of the dump. Table~\ref{tab:data-dumps} shows the months in which Stack Overflow provided database dumps between August 2008 to June 2013. 

\begin{table}[ht!] \scriptsize
\captionsetup{font=scriptsize, labelfont=bf, textfont=bf}
\centering
\caption{shows the months in which Stack Overflow provided database dumps between August 2008 to June 2013.}
\begin{tabular}{lccccc} \hline
\textbf{Months} & \textbf{2009} & \textbf{2010} &  \textbf{2011} &  \textbf{2012} &  \textbf{2013} \\ \hline
January &  & \checkmark  &  \checkmark &   &   \\ 
February &  & \checkmark  &   &   &   \\ 
March &  & \checkmark &   &   &  \checkmark \\ 
April &  & \checkmark  &  \checkmark & \checkmark &   \\ 
May &  & \checkmark &   &   &   \\ 
June &  &  & \checkmark &   &  \checkmark \\ 
July &  & \checkmark   &   &   &   \\ 
August & \checkmark & \checkmark &   & \checkmark &   \\ 
September & \checkmark & \checkmark &  \checkmark &   &  -- \\ 
October & \checkmark & \checkmark  &   &   &  -- \\ 
November & \checkmark & \checkmark  &   &   &  -- \\ 
December & \checkmark &   & \checkmark &   & -- \\ 
\hline
Total & \multicolumn{5}{c}{24 Database Snapshots} \\
Information & \multicolumn{5}{c}{Questions, Answers, Comments, Votes and Badges} \\
\end{tabular}
\label{tab:data-dumps}
\end{table}

We download all the available 24 database dumps for our study. Table~\ref{tab:data-stats} shows the overall statistics of user-generated content on Stack Overflow between August 2008 (inception) to June 2013 (current). The statistics show that Stack Overflow is a very popular programming CQA with 5.1M questions, 9.4M answers and 2.05M registered users.

\begin{table}[ht!] \scriptsize
\captionsetup{font=scriptsize, labelfont=bf, textfont=bf}
\centering
\caption{overall statistics of user-generated content on Stack Overflow between August 2008 -- June 2013}
\begin{tabular}{l|l} \hline
Users & 2.05M (936k askers, 630k answerers) \\ 
Questions & 5.1M (60.22\% with accepted answers) \\ 
Answers & 9.4M  (32.67\% marked as accepted) \\ 
Votes & 42.3M  (70.5\% +ve, 6.7\% favorites)\\ 
Ratio of Answers   & \multirow{2}{*}{1.84} \\
to Questions & ~  \\
\hline
\end{tabular}
\label{tab:data-stats}
\end{table}

However, the database dumps provided by Stack Overflow do not directly contain information about deleted questions. Hence, we analyze the entire 24 database snapshots over $\approx$ 5 years to construct our dataset. Concretely, questions available in the earlier database snapshots (August 2009 -- March 2013) but absent in the most recent snapshot (June 2013) are \emph{deleted} from Stack Overflow. We obtain \textbf{293,289} ($\approx$0.29M) deleted questions between September 2009 and June 2013 by following this procedure. However, we notice a small (but sizeable) percentage of questions in this set which are not deleted but wrongly captured. We eliminate such errors by inspecting HTML pages and obtain a final experimental dataset of \textbf{270,604} ($\approx$0.27M) deleted questions. Table~\ref{tab:delques-data-stats} shows descriptive statistics of user-generated content of deleted questions on Stack Overflow between August 2008 to June 2013.

\begin{table}[ht!] \scriptsize
\captionsetup{font=scriptsize, labelfont=bf, textfont=bf}
\centering
\caption{shows statistics of user-generated content of deleted questions on Stack Overflow between August 2008 to June 2013. The distribution sparklines begin at minimum and end at their corresponding maximum value (log-scale). The distributions are generated using Gaussian kernel density estimates.}
\begin{tabular}{lllllc} 
~ & \textbf{Mean} & \textbf{Median} & \textbf{Min} & \textbf{Max} & \textbf{Distribution} \\ \hline
Questions & \multirow{2}{*}{54,120} & \multirow{2}{*}{49,221} & \multirow{2}{*}{17,514} & \multirow{2}{*}{102,623} &  
\multirow{2}{*}{
\begin{sparkline}{5}
\sparkspike 0.1 0.21914191
\sparkspike 0.3 0.47962932
\sparkspike 0.5 1
\sparkspike 0.7 0.76744005
\sparkspike 0.9 0.1706635
\end{sparkline}
} \\
per year & ~ & ~ & ~ & ~ & ~ \\

Answers & 2.96 & 1.0 & 0.0 & 637.0 & \begin{sparkline}{9}

\spark  0.0 0.774473472397    0.2 0.0651783709582    0.3 0.0448530135865    0.4 0.034563704159    0.5 0.0279924210086    0.6 0.023649305704    0.7 0.0188566804282    0.8 0.0176117198774    0.9 0.0168110179147 /
\end{sparkline} \\

\multirow{2}{*}{Score} & \multirow{2}{*}{0.15} & 2.22 & \multirow{2}{*}{-56.0} & \multirow{2}{*}{695.0} & \multirow{2}{*}{\begin{sparkline}{9}
\spark -0.0833333333333 0.0    -0.00333333333333 0.737098237201    0 0.999999999997    0.0833333333333 0.178994160055    0.166666666667 0.0    0.5 0.0    0.666666666667 0.0    0.833333333333 0.0    1.0 0.0 /
\end{sparkline}} \\
~ & ~ & x$10^{-16}$ & ~ & ~ & ~ \\

Views & 221.65 & 80.0 & 1.0 & 296,466 & \begin{sparkline}{9}
\spark 0.1 0.00711020822753    0.2 0.227018791265    0.3 0.652107668867    0.4 0.504824784154    0.5 0.0228542407313    0.6 0.00304723209751    0.7 0.000507872016252    0.8 0.000507872016252    0.9 0.000507872016252 /
\end{sparkline} \\

Favorites & 3.59 & 1.0 & 0.0 & 1530.0 & \begin{sparkline}{9}

\spark  0.0 0.774473472397    0.2 0.0651783709582    0.3 0.0448530135865    0.4 0.034563704159    0.5 0.0279924210086    0.6 0.023649305704    0.7 0.0188566804282    0.8 0.0176117198774    0.9 0.0168110179147 /
\end{sparkline} \\

Comments & 2.26 & 1.0 & 0.0 & 77.0 & \begin{sparkline}{9}
\spark 0.0 0.730521149493 0.1 0.5294926634662 0.2 0.000651 0.3 0.0001 0.4 0.0001 0.5 0.000001 0.6 0.000001 0.7 0.000001 0.8 0.000001 0.9 0.000001  /
\end{sparkline} \\

\hline
\end{tabular}
\label{tab:delques-data-stats}
\end{table}

We would like to point out that our experimental dataset contains a subset of all deleted questions on Stack Overflow. This limitation is due to the sporadic sharing of database dumps by Stack Overflow (and not due to the procedure we use to find deleted questions). Questions deleted between two data snapshots would not be captured in our experimental dataset. However, it must be noted that our dataset contains the maximum possible deleted questions which can be obtained given the publicly available Stack Overflow database snapshots. We also make our experimental dataset publicly available for research purposes under the \emph{Creative Commons Attribute-ShareAlike}.\footnote{\scriptsize{\url{http://correa.in/datasets.html}}}

\subsection{Increase in Deleted Questions Over Time}

\begin{figure}[ht!]
\captionsetup{font=scriptsize, labelfont=bf, textfont=bf}
\centering
\includegraphics[width=\linewidth]{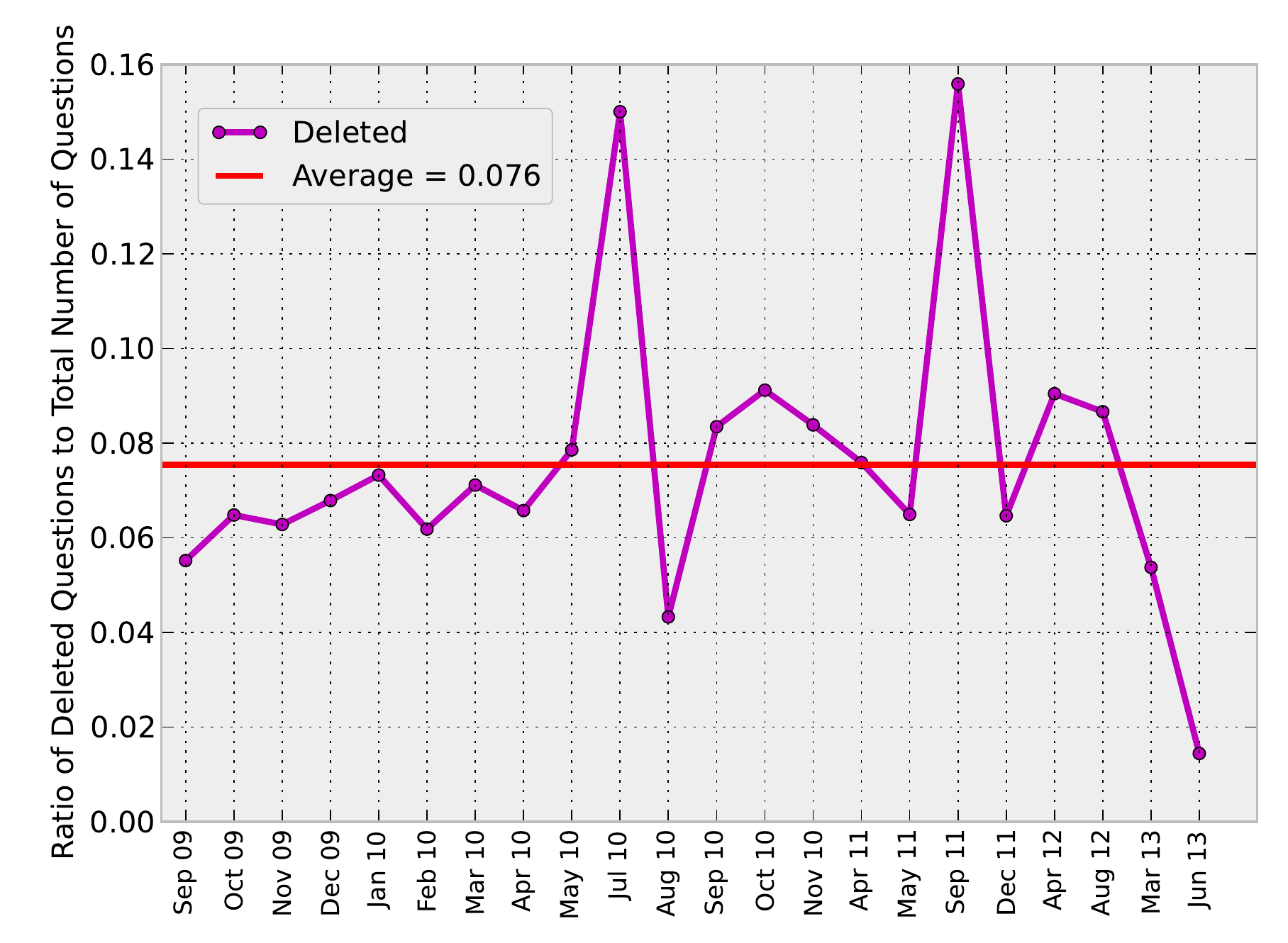}
\caption{shows the ratio of deleted questions to total questions in each month over a 49-month period between September 2009 and June 2013.}
\label{fig:del-ques-temporal}
\end{figure}

We now perform a temporal trend analysis of deleted questions on Stack Overflow. Figure~\ref{fig:del-ques-temporal} shows the ratio of deleted questions to total questions in each month over a 49-month period between September 2009 and June 2013. We would like to recollect that the first database snapshot provided by Stack Overflow is on August 2009 and hence, we do not have information of deleted questions prior to this snapshot. We observe that on an average $\approx$8\% of questions are deleted on Stack Overflow. We also notice an anomalous increase in percentage of deleted questions $\approx$15\% after the month of May (2010 2011). We posit that these abrupt rises could be due to periodic \emph{Deletion Question Audits} conducted by the Stack Overflow community around the month of May~\cite{:2010kx}. 

\begin{figure}[ht!]
\captionsetup{font=scriptsize, labelfont=bf, textfont=bf}
\centering
\includegraphics[width=\linewidth]{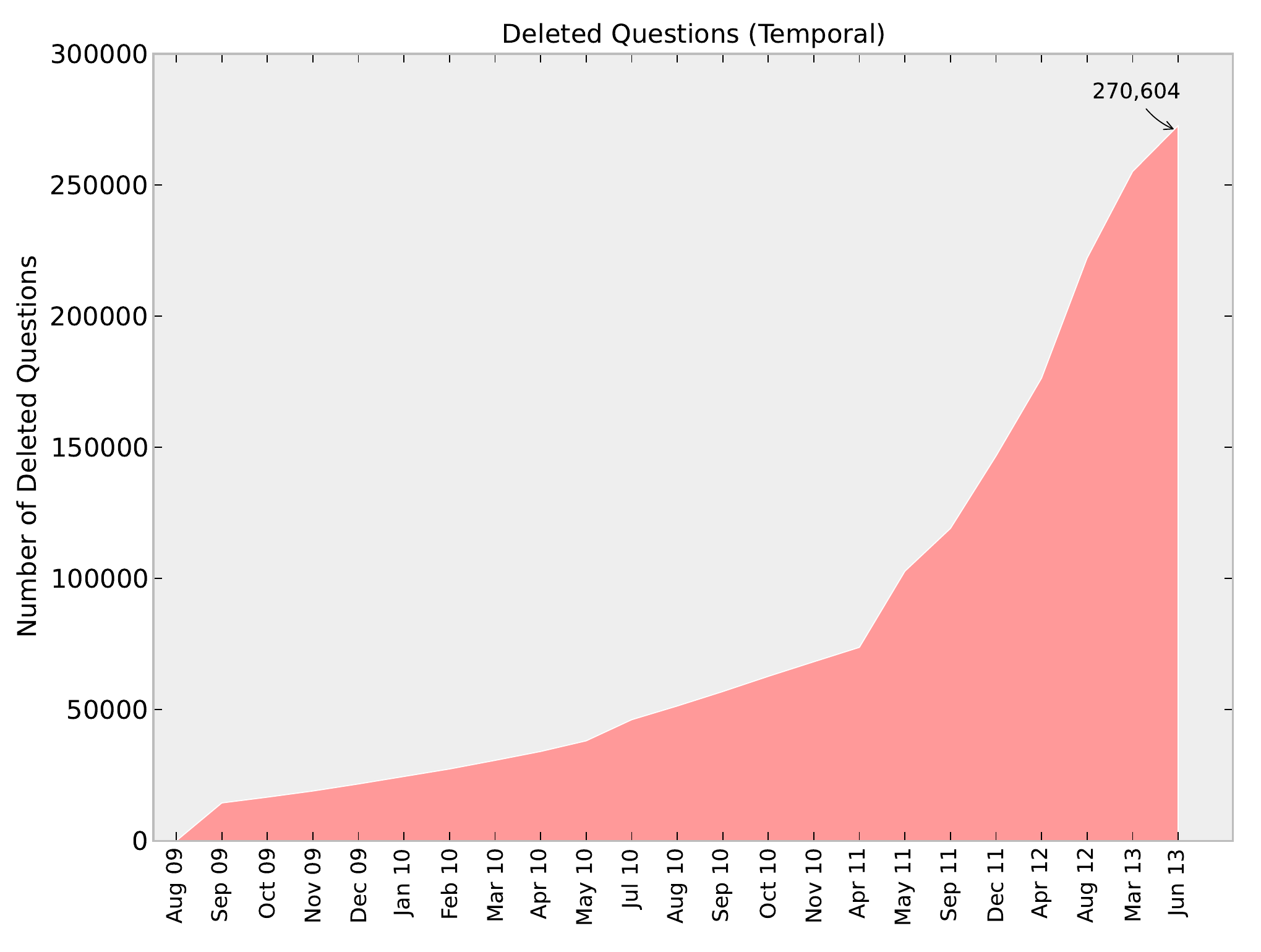}
\caption{shows the cumulative distribution of deleted questions over a 49-month period between September 2009 and June 2013. We notice a sharp rise in the number of deleted questions over time.}
\label{fig:del-ques-temporal-cml}
\end{figure}

Figure~\ref{fig:del-ques-temporal-cml} shows the cumulative distribution area chart of deleted questions over a 49-month period between September 2009 and June 2013. The area chart depicts that there is a sharp increase in the total number of deleted questions over time (even with an average of $\approx$8\%). We notice a particularly steep increase after May 2011. Therefore, despite the presence of comprehensible and explicit question posting guidelines -- Stack Overflow receives a high number of extremely poor quality questions which are not fit to exist on its website. Hence, it is important to perform a longitudinal study about deleted questions on Stack Overflow.

\subsection{Community Takes Long Time to Detect but Swift Action by Moderators}

\begin{figure}[ht!]
\captionsetup{font=scriptsize, labelfont=bf, textfont=bf}
\centering
\includegraphics[width=\linewidth]{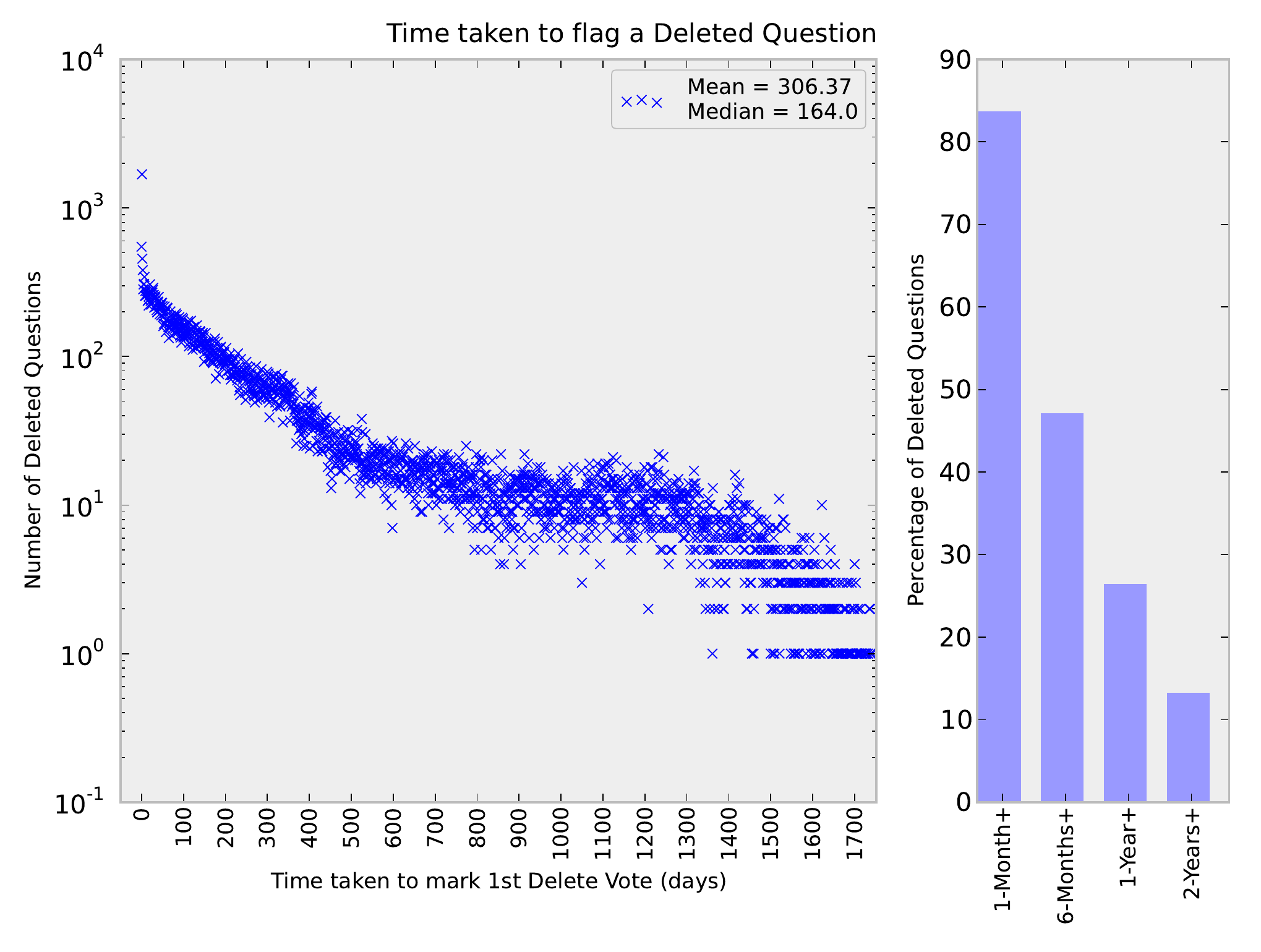}
\caption{shows the distribution of time taken to receive the first delete vote on a deleted question on Stack Overflow.}
\label{fig:del_flag_distr}
\end{figure}

Stack Overflow delineates an elaborate procedure to delete a question. We recall that experienced community members viz. users with 10,000+ reputation points can cast `delete votes' to delete a question. We analyze these `delete vote' patterns to gain insights into the community participation dynamics on poor quality questions on Stack Overflow. Hence in this section, we restrict our analysis to \textbf{62,949} deleted questions which have received `deleted votes'. Figure~\ref{fig:del_flag_distr} shows the distribution of time taken to receive the first `delete vote' on a deleted question on Stack Overflow. We see that $\approx$80\% questions take at least 1 month(or more) to receive its first `delete vote' while $\approx$50\% of the questions take 6 months(or more). Overall, 50 percentile of the questions take $>$164 days to receive their first `delete vote'. Therefore, majority of deleted questions take a significant amount of time to receive its first delete vote.

\begin{figure}[ht!]
\captionsetup{font=scriptsize, labelfont=bf, textfont=bf}
\centering
\includegraphics[width=\linewidth]{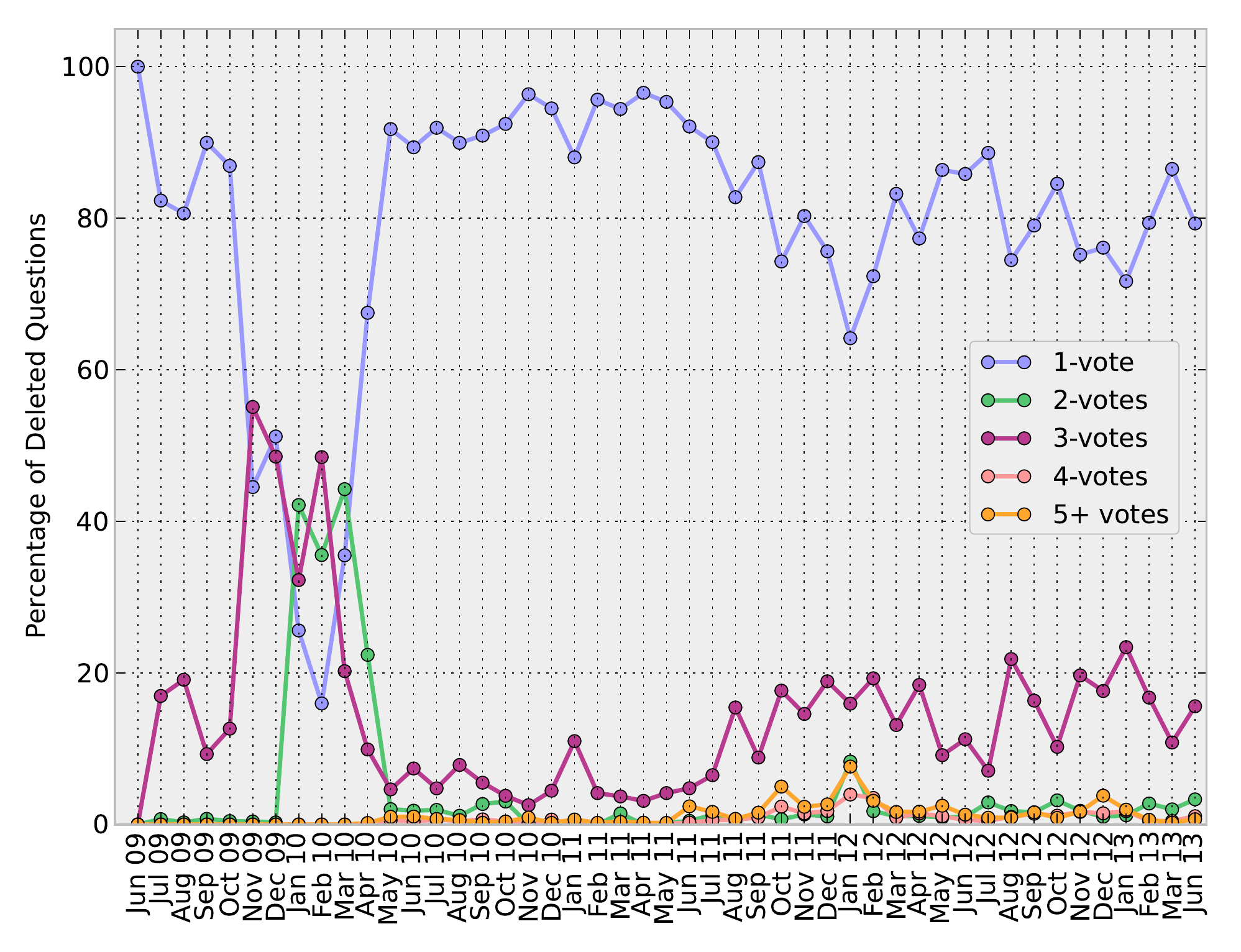}
\caption{shows the temporal distribution of `delete votes' on deleted questions in Stack Overflow between June 2009 and June 2013.}
\label{fig:del-vote-temporal}
\end{figure}

We now analyze the distribution of `deleted votes' in our dataset. Table~\ref{tab:del-vote-distr} shows the `deleted vote' distribution on deleted questions posted and Figure~\ref{fig:del-vote-temporal} shows the temporal distribution of `delete votes' on deleted questions. We observe that $\approx$80\% questions receive 1 `delete vote' and $\approx$14\% questions receive 3 `delete votes' before they are deleted. This shows that once a question receives the required number of `delete vote'(s), moderators move quickly to take appropriate action. The moderators are driven by the Stack Overflow motto to keep a low signal-to-noise ratio in order to maintain high content quality.  We also note that $\approx$23\% of questions in our experimental dataset have received `delete votes' i.e. questions which are voted to be deleted by experienced users. Therefore, 75\% of questions in our dataset are never voted for deletion by the community. This reveals that the major duty to detect extremely poor quality questions is on the encumbrance of the moderator. This, coupled with the `delete vote' pattern-action behavior (relatively long time but swift action) motivates the impending need to have a predictive system for the benefit of the moderators on Stack Overflow.

\begin{table}[ht!] \scriptsize
\captionsetup{font=scriptsize, labelfont=bf, textfont=bf}
\centering
\caption{shows the `deleted vote' distribution on deleted questions posted between June 2009 and June 2013. }
\begin{tabular}{c|c} \hline
\textbf{Votes} & \textbf{Deleted Questions} \\ \hline
1-vote & 50,012 (79.45\%)  \\ 
2-votes & 2,736 (4.35 \%)  \\ 
3-votes & 9,009 (14.3\%)   \\
4-votes & 463 (0.74\%)  \\
5+-votes & 729 (1.16\%)  \\ \hline
Total & 62,949  \\ 
\hline
\end{tabular}
\label{tab:del-vote-distr}
\end{table}

\subsection{Authors Delete Questions to Salvage Reputation}

\begin{figure}[ht!]
\captionsetup{font=scriptsize, labelfont=bf, textfont=bf}
\centering
\includegraphics[width=\linewidth]{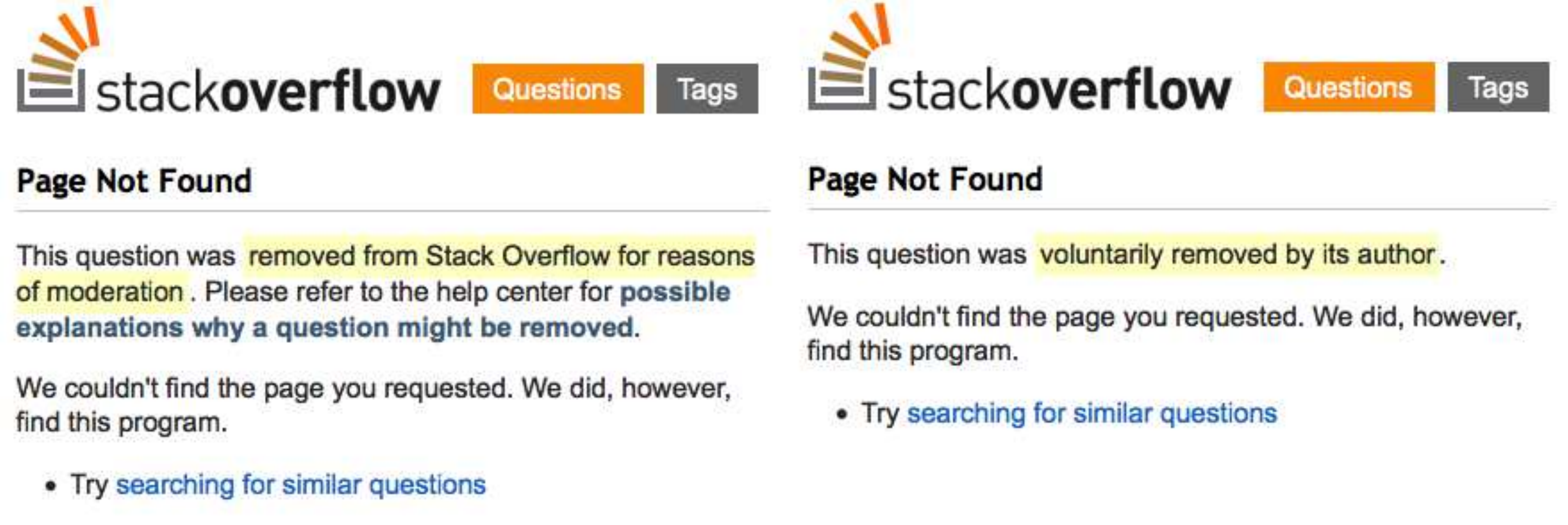}
\caption{shows the web page screenshots of two questions deleted by moderator (left) and author (right) on Stack Overflow.}
\label{fig:del-ques-example}
\end{figure}

We recall that a question on Stack Overflow can either be deleted by the author of the question or by a moderator. However, this information is not directly available in the publicly available data dumps provide by Stack Overflow. We also recall that questions on Stack Overflow are not digitally deleted i.e. they are hidden from the site and do not appear in search results. We notice that this information about question deletion initiator (question author or moderator) is available on the unique web address of the question. Figure~\ref{fig:del-ques-example} shows the web page screenshots of -- (i) question deleted by moderator (left) and (ii) question deleted by author (right). 

\begin{figure}[ht!]
\captionsetup{font=scriptsize, labelfont=bf, textfont=bf}
\centering
\includegraphics[scale=0.5]{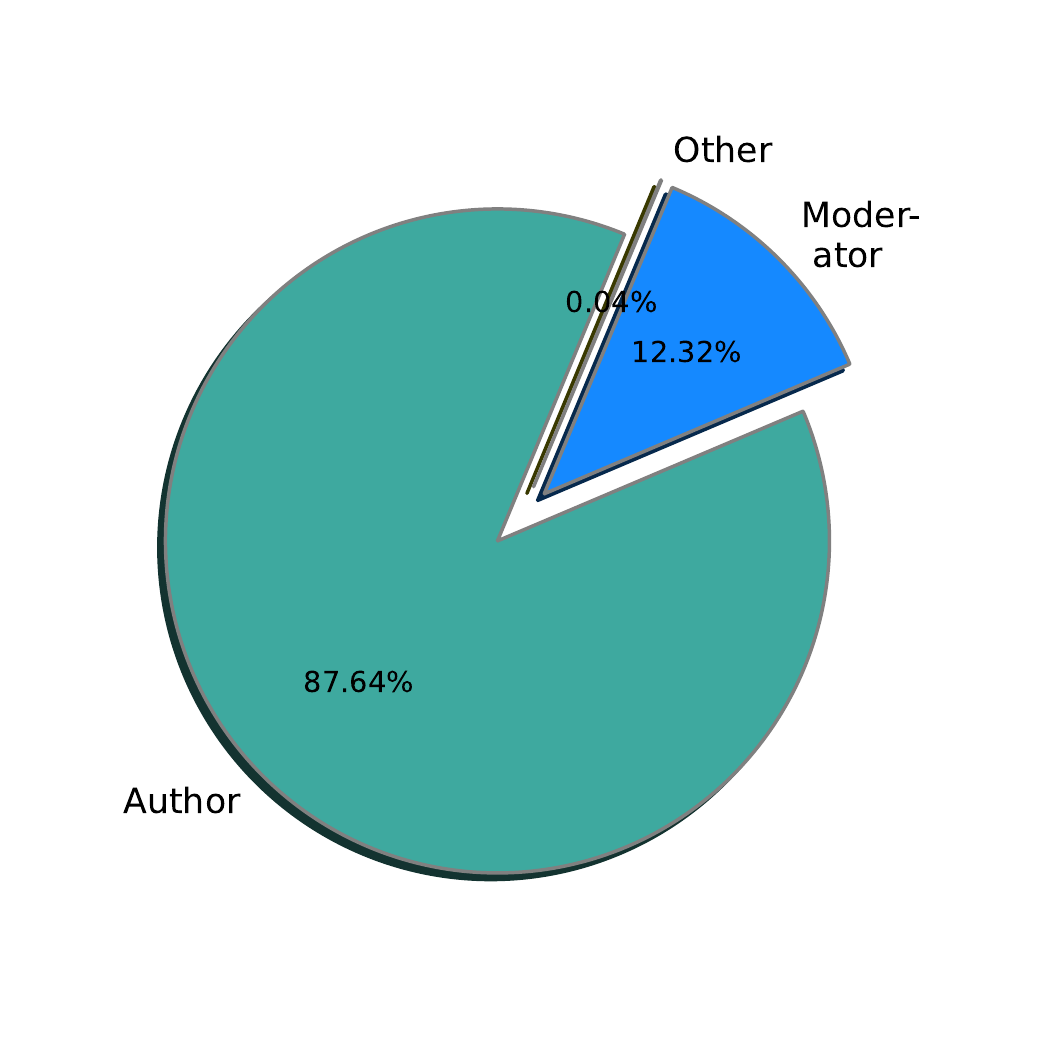}
\caption{shows the distribution of question deletion initiator (moderator or author) on Stack Overflow}
\label{fig:del-reasonl}
\end{figure}

We download the unique web pages of deleted questions in our experimental dataset and employ a regular expression to extract this information. Figure~\ref{fig:del-reasonl} shows the distribution of question deletion initiator (moderator or author) on Stack Overflow. We notice that $\approx$87\% (237,163) of questions are deleted by a moderator while $\approx$12\% (33,330) or 1 out of 8 questions are deleted by the question author. A negligible percentage (0.04\%) of questions (111) do not belong to either category. This may be a bug in the data dump and hence, we ignore these questions in this section.

\begin{figure}[ht!]
\captionsetup{font=scriptsize, labelfont=bf, textfont=bf}
\centering
\includegraphics[width=\linewidth]{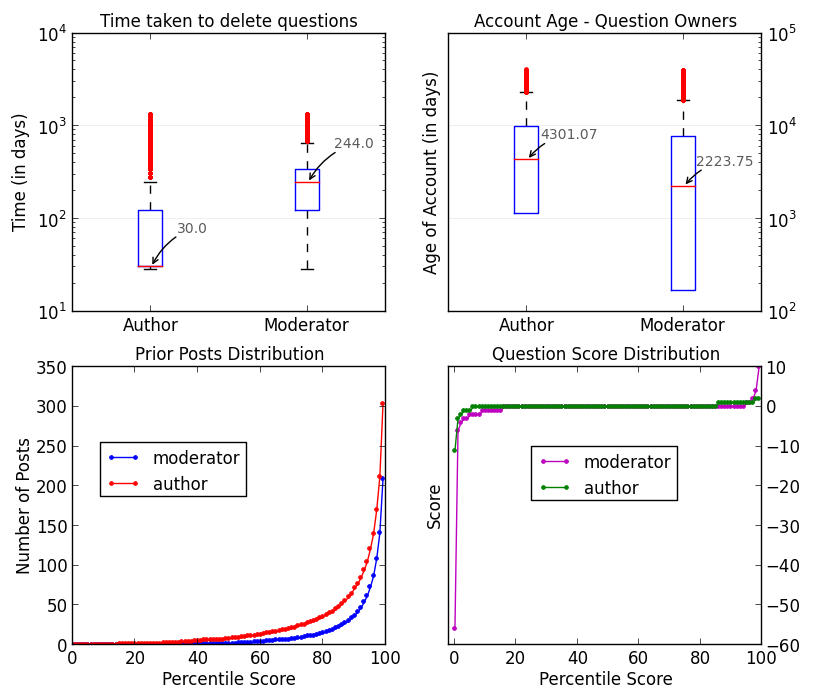}
\caption{shows the distributions of -- (top-left) time taken to delete a question via box-and-whisker plot, (top-right) account age of question owner via box-and-whisker plot, (bottom-left) posts prior to deleted question creation via percentile plot and (bottom-right) deleted question score via percentile plot for author and moderator deleted questions.}
\label{fig:mod_auth_comparison}
\end{figure}

We now analyze multiple question attributes based on the question deletion initiator (author or moderator) and make observations. Figure~\ref{fig:mod_auth_comparison} (Top-Bottom, Left-Right) shows the distributions of -- (i) time taken to delete a question (box-and-whisker plot), (ii) account age of question owner (box-and-whisker plot), (iii) posts prior to deleted question creation (percentile plot) and (iv) deleted question score (percentile plot) for author and moderator deleted questions. The Top-Left box-and-whisker plot reveals that the distribution of time taken to delete a question by the question author is relatively lower than when done by a moderator. The Top-Right box-and-whisker plot shows that the question owner age of account is relatively higher of author deleted questions than moderator deleted questions. This points out that question owners of author-deleted questions are more experienced on the website(in terms of time) than question owners of moderator-deleted questions. The Bottom-Left percentile plot shows that prior posts  -- posts made by the question owner prior to the time of deleted question creation -- of author-deleted questions is higher than those deleted by a moderator. The Bottom-Right percentile show shows that question scores of author-deleted questions are higher than those of moderator-deleted questions at very low ($<$20) and negative question scores. 

The above observations show that despite being less experienced on the website, question owners of author-deleted questions have more prior posts and have higher question scores on deleted questions than those of moderator-deleted questions. On the other hand, author-deleted questions take lesser time than those of moderator-deleted questions. We argue that question owners of author-deleted questions exhibit such a behavior as they want to maintain a healthy reputation earned on Stack Overflow~\cite{bosu2013building}. The owners of author-deleted questions observe that their questions are attracting down votes  which affects their \emph{reputation}. In order to stem further decrease reputation points, question owners see deletion of question as a quick fix and therefore, proceed to delete the posted question in an attempt to salvage their reputation.

\subsection{Question Quality Pyramidal Structure}

Questions on Stack Overflow are marked `closed' if they are deemed unfit for the question-answer format on Stack Overflow and indicate low quality. A question can be marked as `closed' due to five reasons -- \emph{duplicate}, \emph{subjective}, \emph{off topic}, \emph{too localized} or \emph{not a real question}~\cite{Correa:2013vn}. We find that there are \textbf{254,446} (0.25M) `closed' questions on Stack Overflow between August 2008 June 2013. In this section, in addition to our experimental dataset we utilize this data to analyze question quality indicators for deleted and `closed' questions. We also make observations about the relative quality of deleted questions in context to `closed' questions on Stack Overflow. 

\paragraph{\textbf{Community Value}}

\begin{figure}[ht!]
\captionsetup{font=scriptsize, labelfont=bf, textfont=bf}
\centering
\includegraphics[width=\linewidth]{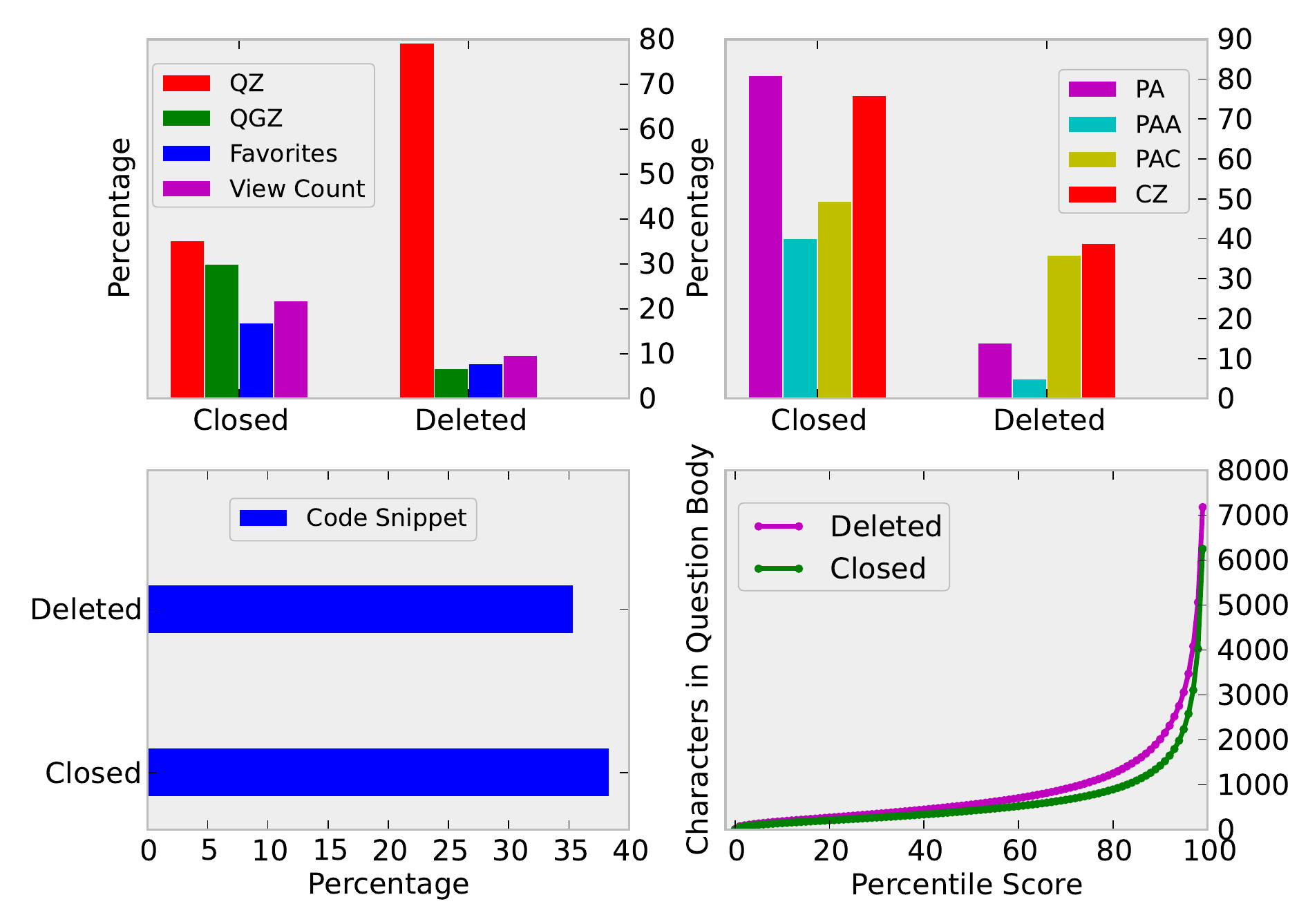}
\caption{shows the quantities of question quality indicators for `closed' and deleted questions on Stack Overflow.(QZ=Percentage of Questions with Zero Score, QGZ=Percentage of Questions with Greater than Zero Score, PA=Percentage of Answers, PAA=Percentage of Accepted Answers, PAC=Percentage of Accepted Answers given that a question has an answer, CZ=Percentage of Comments with Zero Scores)}
\label{fig:cloq_del_quality}
\end{figure}

Figure~\ref{fig:cloq_del_quality} shows various quantities of question quality indicators for `closed' and deleted questions on Stack Overflow.  We see that deleted questions have higher percentage of questions with zero \emph{score} than `closed' questions. A question \emph{score} is the net worth of the usefulness of a question as determined by the Stack Overflow community. $\approx$80\% of deleted questions have a zero \emph{score} which indicates that most deleted questions are of very little worth to the community. We also observe that in comparison to `closed' questions -- deleted questions have a lower percentage of question with greater than zero \emph{score}, \emph{favorite} votes and \emph{view counts}. A \emph{favorite} vote is equivalent to an explicit expression of interest or subscription on a question. Only $\approx$5-6\% of deleted questions attract a positive question \emph{score} or \emph{favorite} votes. This indicates that deleted questions are generally of very little worth and interest to the Stack Overflow community. In addition, $\approx$10\% of deleted questions have $>$0 \emph{view counts} which is twice as lower than that of closed questions ($\approx$23\%). We also find that deleted questions have lesser percentage of code snippets but slightly higher number of characters in their body than `closed' questions. Nonetheless, we observe that deleted questions fare inferior on most indicators in comparison to `closed' questions. This signifies that deleted questions are extremely poor in quality with respect to closed questions (which themselves are low quality).

\paragraph{\textbf{Question Topics}}

\begin{figure}[ht!]
\captionsetup{font=scriptsize, labelfont=bf, textfont=bf}
\centering
\includegraphics[scale=0.17]{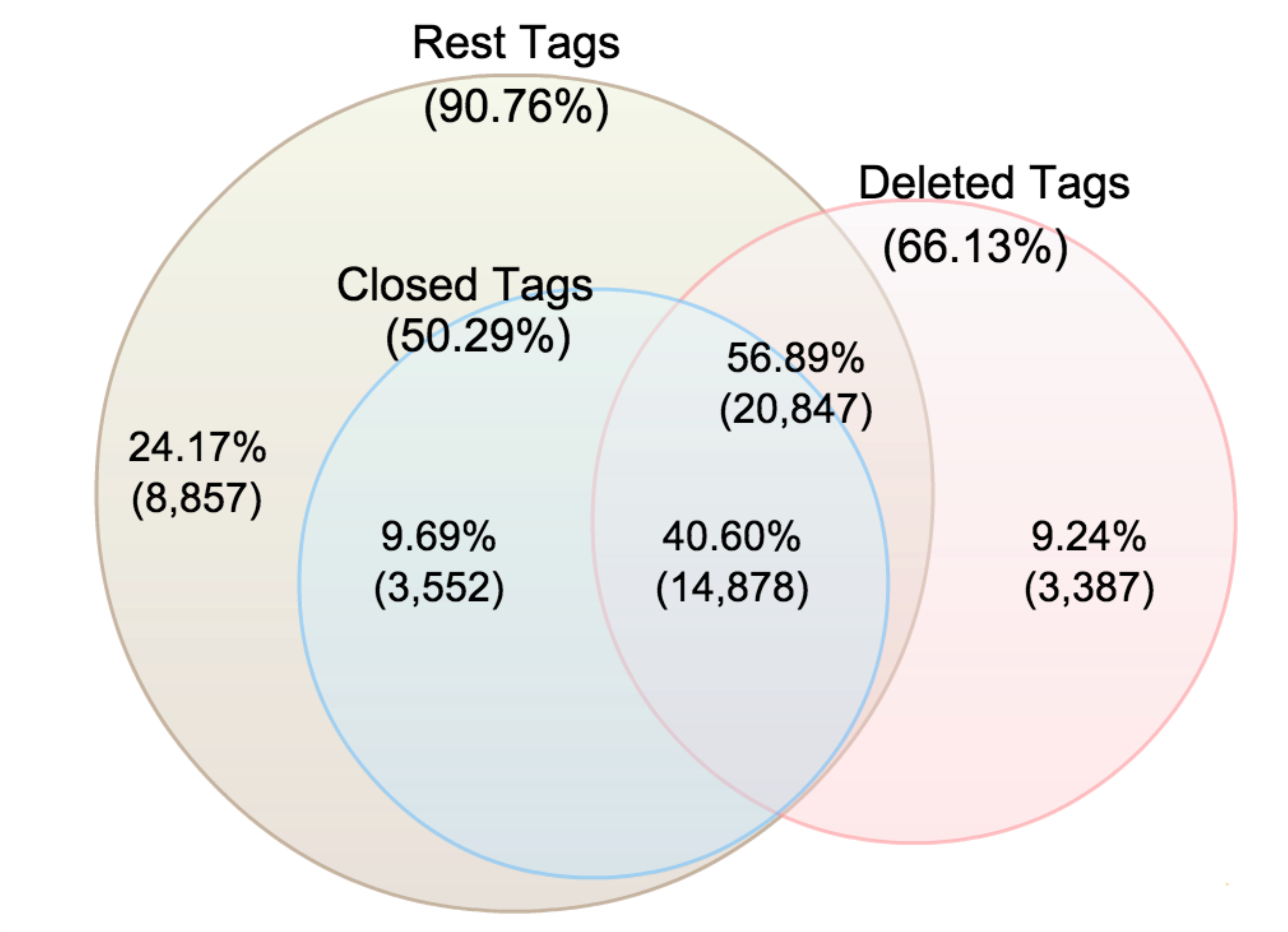}
\caption{ shows the venn diagram of tag distributions of questions on Stack Overflow.}
\label{fig:cloq_del_venn}
\end{figure}

Stack Overflow questions contain user supplied \emph{tags} which indicate the topic of the question. We analyze the tag distribution of closed and deleted questions and compare them to the overall tag distribution on Stack Overflow. There are a total of \textbf{36,643} tags on all questions in Stack Overflow. Figure~\ref{fig:cloq_del_venn} shows the venn diagram of tag distributions of questions on Stack Overflow. 

\begin{figure}[ht!]
\captionsetup{font=scriptsize, labelfont=bf, textfont=bf}
\centering
\includegraphics[width=\linewidth]{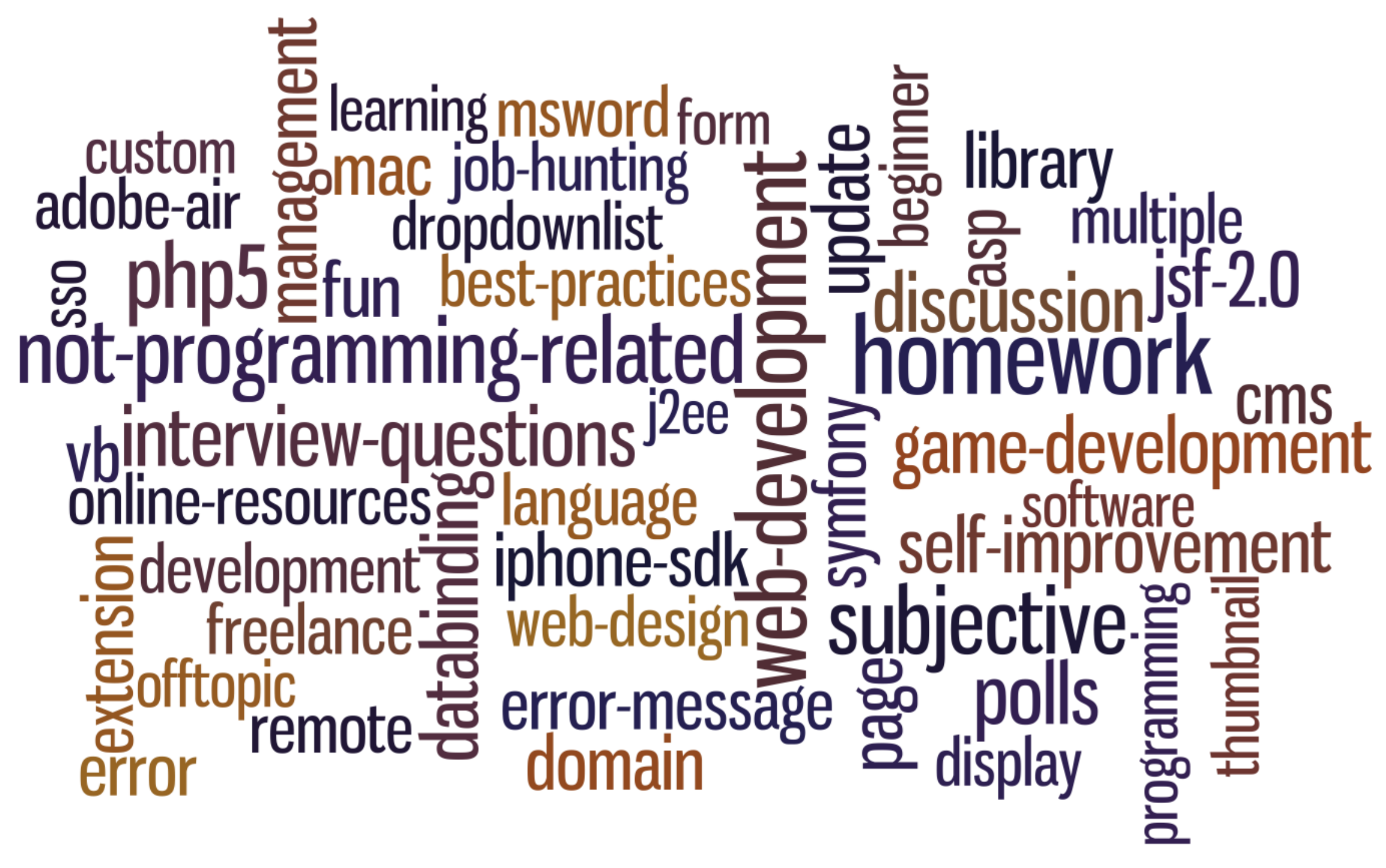}
\caption{shows the word cloud of the top-50 tags that occur on deleted questions. 9.24\% of tags are unique to deleted questions on Stack Overflow.}
\label{fig:tagcloud_deleted_tags}
\end{figure}

We see that tags on `closed' questions are a subset of the overall tags which occur in regular questions. In contrast, an appreciable number of tags on deleted questions ($\approx$10\%) are not found either in `closed' or regular questions. We extract such tags found in deleted questions for further analysis. The topics of `closed' questions are relevant to the community despite the questions themselves being unfit for the Stack Overflow format. However, some of the topics on deleted questions are extremely off topic to the interests of the community. Figure~\ref{fig:tagcloud_deleted_tags} shows the word cloud of the top-50 tags that occur on deleted questions. We find a very high presence of low quality tags like \emph{homework}, \emph{job-hunting} and \emph{polls} on deleted questions. These topics also show that deleted questions are of extremely poor quality.

\paragraph{\textbf{Effort to Improve Question Quality}}

The goal of Stack Overflow is to be a comprehensive knowledge base for programming related topics. Therefore, a question on Stack Overflow may be edited by privileged users (experienced community members and moderators) to maintain content quality. A question on Stack Overflow has three major sections which can be edited -- title, body and tags. In addition, users without sufficient privileges can \emph{suggest edits} to questions. These suggestions are then reviewed by privileged users and are brought into effect at their discretion. In this part, we analyze edit histories of deleted questions on Stack Overflow. Table~\ref{tab:del-hist-distr} shows four edit types (edit tags, edit body, edit tile and suggested edits) for deleted questions in our experimental dataset. We see that $\approx$93\% of deleted questions receive at least one form of edit. Also, moderator-deleted questions receive more edits than author-deleted questions. 

\begin{table}[ht!] \scriptsize
\captionsetup{font=scriptsize, labelfont=bf, textfont=bf}
\centering
\caption{shows edit details of deleted questions on Stack Overflow}
\begin{tabular}{l|l|l|l} \hline
~ & \textbf{Moderator} & \textbf{Author} & \textbf{Total} \\ \hline
Deleted Questions & 237,163 & 33,330 & 270,493 \\ 

\multirow{2}{*}{Question Content Edited} & 226,286 & 26,461 & 252,747 \\ 
~ & (95.41\%) & (79.39 \%) & (93.44\%) \\

\multirow{2}{*}{Question `Closed'} & 34,209  & 2.167 & 36,376 \\
~ & (15.12\%) & (8.19\%) & (14.38\%) \\


\hline
\end{tabular}
\label{tab:del-hist-distr}
\end{table}

\begin{figure}[ht!]
\captionsetup{font=scriptsize, labelfont=bf, textfont=bf}
\centering
\includegraphics[width=\linewidth]{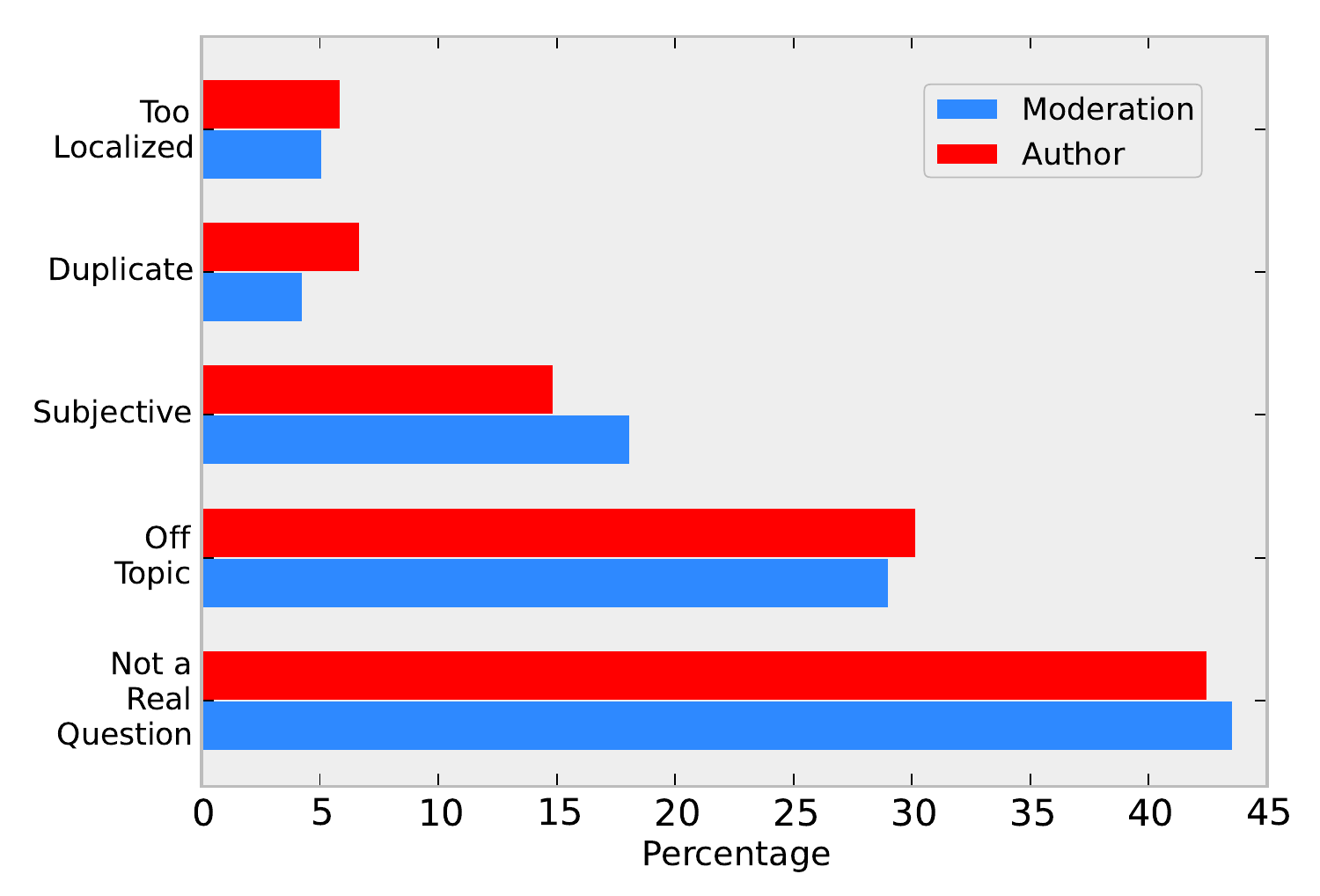}
\caption{shows the distribution of questions marked as `closed' on five reasons for author and moderator deleted questions.}
\label{fig:cloq_posthistory}
\end{figure}

We also notice that 14.38\% of deleted questions were marked as `closed' before they were deleted -- a higher percentage of moderator-deleted questions were marked as `closed' than author-deleted questions. Figure~\ref{fig:cloq_posthistory} shows the distribution of questions marked as `closed' due to five reasons for both -- author and moderator deleted questions. We see that a higher percentage of author-deleted questions are marked as \emph{too localized}, \emph{duplicate} and \emph{off topic}. However, a higher percentage of moderator-deleted questions are marked as \emph{subjective} and \emph{not a real question}.

\begin{figure}[ht!]
\captionsetup{font=scriptsize, labelfont=bf, textfont=bf}
\centering
\includegraphics[width=\linewidth]{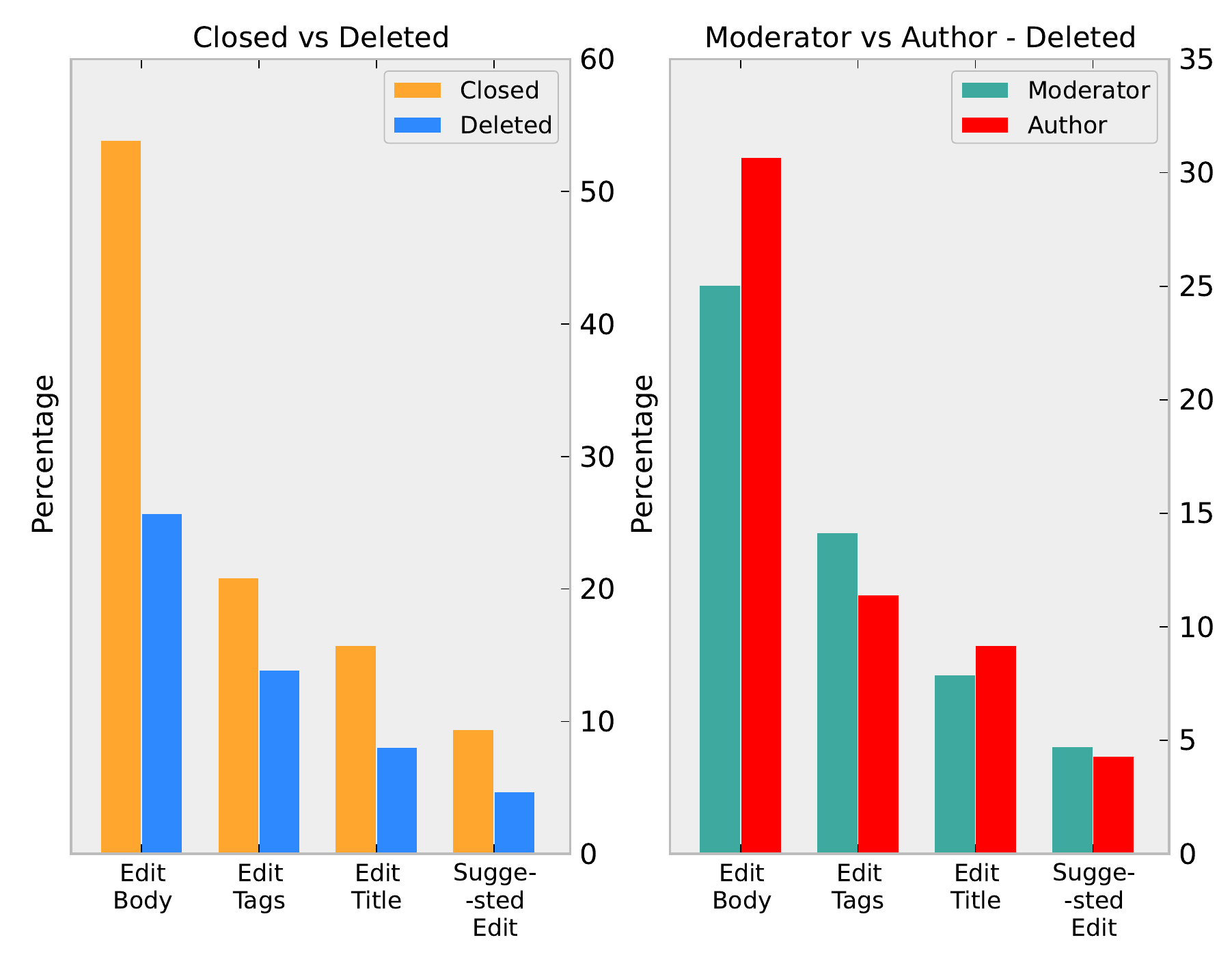}
\caption{shows the distribution of edit question history on (left) `closed' versus deleted and (right) moderator versus author deleted questions.}
\label{fig:del_posthistory}
\end{figure}

Figure~\ref{fig:del_posthistory} shows the distribution of percentages of questions on various edit categories for (left) `closed' versus deleted and (right) moderator versus author deleted questions. We observe that `closed' questions receive a higher percentage of edits than deleted questions. This signifies that the community puts a greater effort to edit and improve `closed' questions than it does for deleted questions. We also see that core content of a question (title and body) for author-deleted questions receive a higher percentage of edits than moderator-deleted questions. This shows that author-deleted questions are inferior in quality than moderator-deleted questions and require more work to improve their content. 

\paragraph{\textbf{Quality Pyramid}}

\begin{figure}[ht!]
\captionsetup{font=scriptsize, labelfont=bf, textfont=bf}
\centering
\includegraphics[scale=0.17]{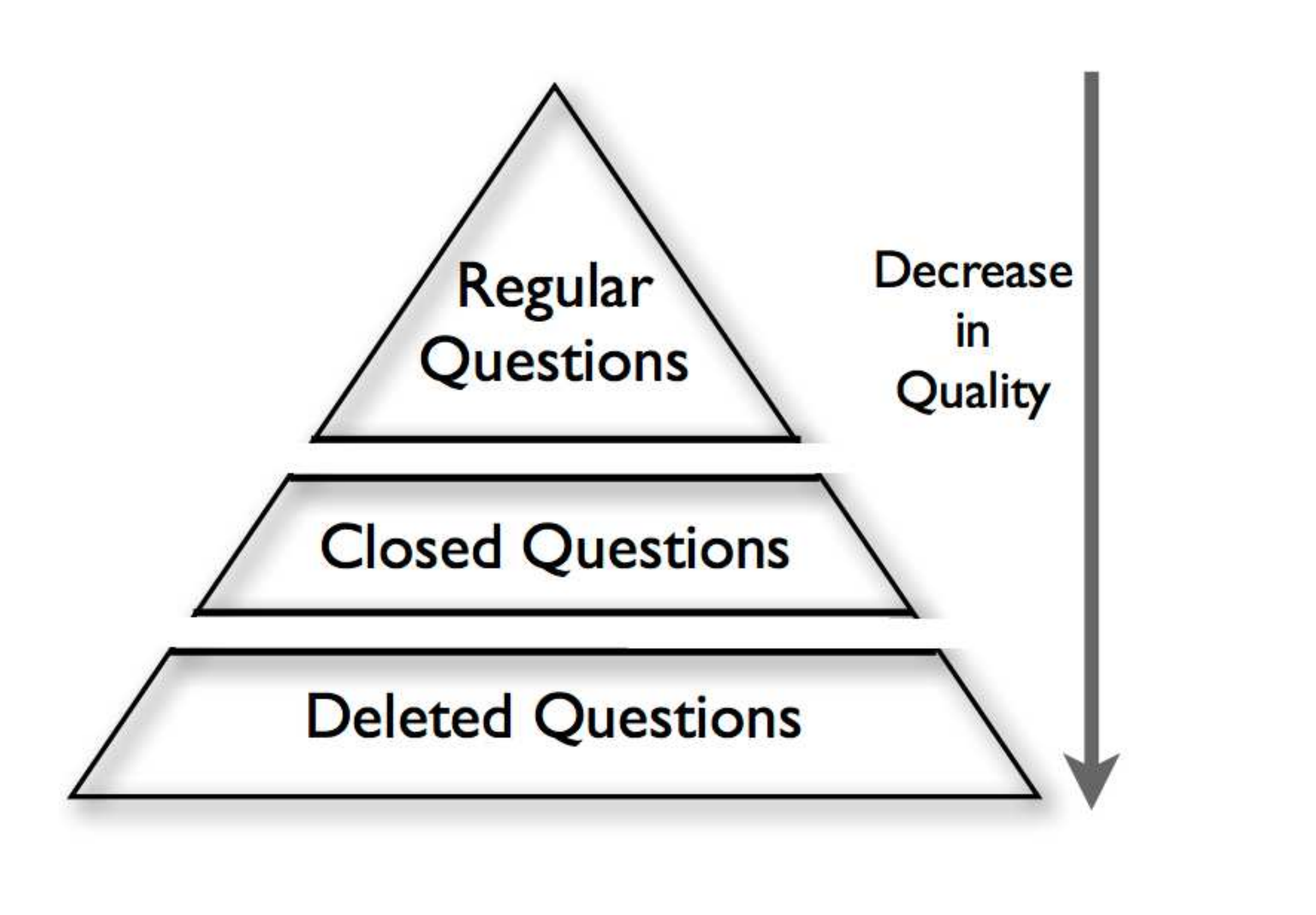}
\caption{shows the underlying question quality pyramid structure on Stack Overflow.}
\label{fig:quality-pyramid}
\end{figure}

`Closed' questions are questions which are deemed unfit for the Stack Overflow format. A `closed' question is low quality but has the potential to be improved upon. On the other hand, deleted questions are very poor in nature and beyond improvement. Prior analysis in this section revealed that deleted questions fare extremely poor on multiple community value quality indicators. We also find that some topics of deleted questions are entirely irrelevant to the Stack Overflow website. In addition, we see that the community puts in more effort to improve a `closed' question than it does for a deleted question. These findings reveal that question quality on Stack Overflow has a pyramidal structure -- regular questions lie at the top of the pyramid (good quality), followed by `closed' questions (bad quality) and deleted questions are placed at the bottom of the pyramid (extremely poor quality). Figure~\ref{fig:quality-pyramid} shows this underlying question quality pyramid structure on Stack Overflow. 

\subsection{Accidental Question Deletion}

Stack Overflow provides a procedure to \emph{undelete} a deleted question. The similar voting procedure to that of deletion is followed to \emph{undelete} a question. Table~\ref{tab:undel-ex-distr} shows some examples of undeleted questions on Stack Overflow.

\begin{table}[ht!] \scriptsize
\captionsetup{font=scriptsize, labelfont=bf, textfont=bf}
\centering
\caption{shows examples of Undeleted Questions on Stack Overflow. }
\begin{tabular}{l|l|l} \hline
\textbf{Id} & \textbf{Title} & \textbf{Score}  \\ \hline
145 & Compressing / Decompressing Folders   & 10 \\
~ & \& Files in C\#? & ~ \\
249 &Accessing a remote form in php  & 15  \\
5235643 & Colon(:) and number in filename in Visio & 1 \\
5767118 & Facebook text field detection & 0 \\
\hline
\end{tabular}
\label{tab:undel-ex-distr}
\end{table}

\begin{figure}[ht!]
\captionsetup{font=scriptsize, labelfont=bf, textfont=bf}
\centering
\includegraphics[scale=0.4]{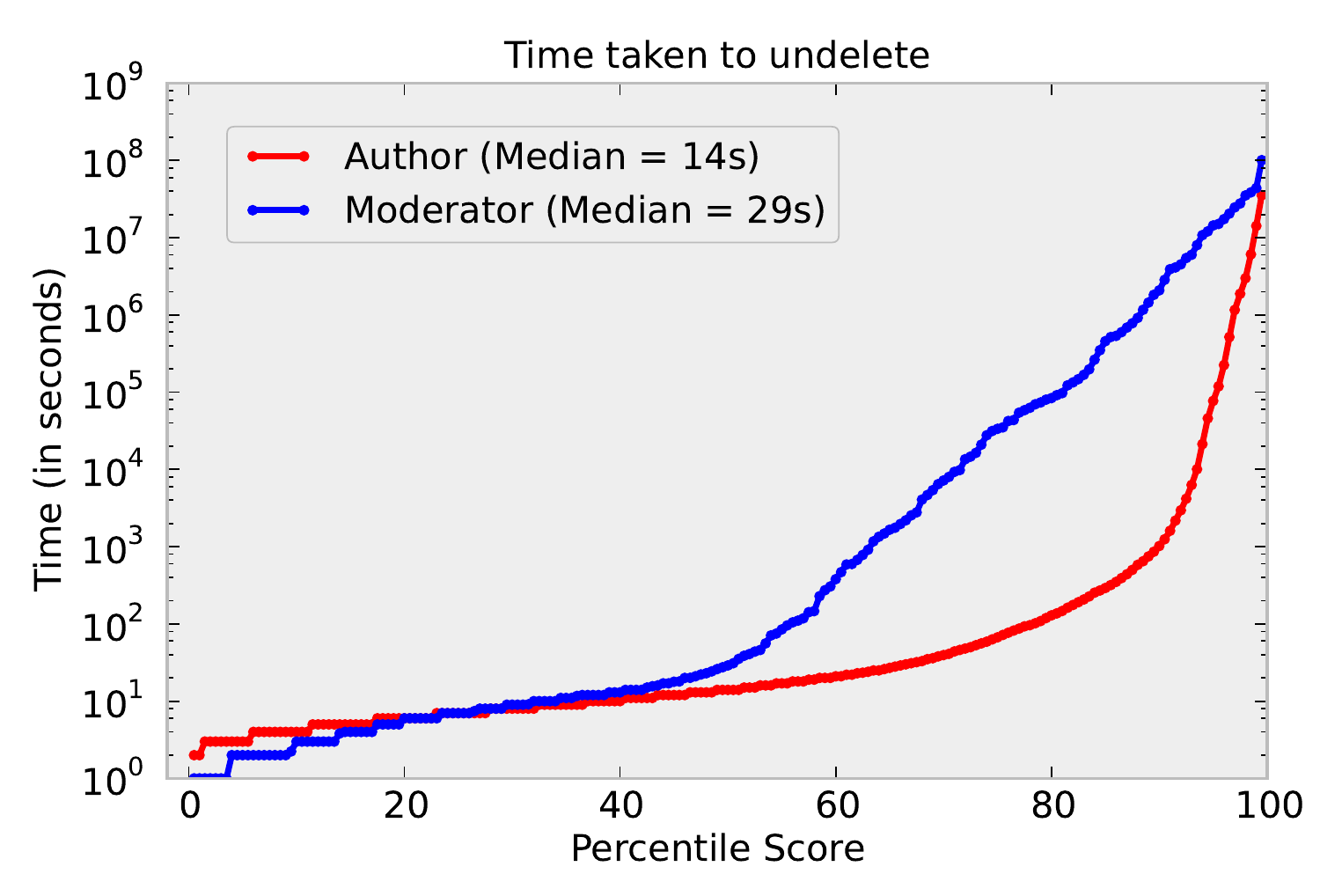}
\caption{shows the percentile plot of time taken to \emph{undelete} a question from the time of deletion of author-deleted and moderator-deleted questions.}
\label{fig:undeleted_time}
\end{figure}

We find a total of \textbf{9,350} \emph{undeleted} questions on Stack Overflow. \textbf{8,536} of these total questions were originally deleted by the question author while \textbf{814} questions were deleted by a moderator. We now analyze the time taken to \emph{undelete} a question from the time of deletion. Figure~\ref{fig:undeleted_time} shows the percentile plot of time taken to \emph{undelete} a question from the time of deletion of author-deleted and moderator-deleted questions. We find that most author-deleted questions are \emph{undeleted} within 2 minutes of its deletion. We attribute this peculiar behavior to accidental deletion. The question author accidentally deletes her question but on realization of her mistake, she \emph{undeletes} the question. We observe that moderator-deleted questions take a relatively longer time to \emph{undelete}. This lag may be due to the time required for the \emph{undelete} voting procedure as defined by the Stack Overflow community guidelines. The guidelines specify that a deleted question must receive a minimum of 3 \emph{undelete} votes to be \emph{undeleted}. This leads to an increase in time taken \emph{undelete} moderator-deleted questions.

\begin{figure}[ht!]
\captionsetup{font=scriptsize, labelfont=bf, textfont=bf}
\centering
\includegraphics[width=\linewidth]{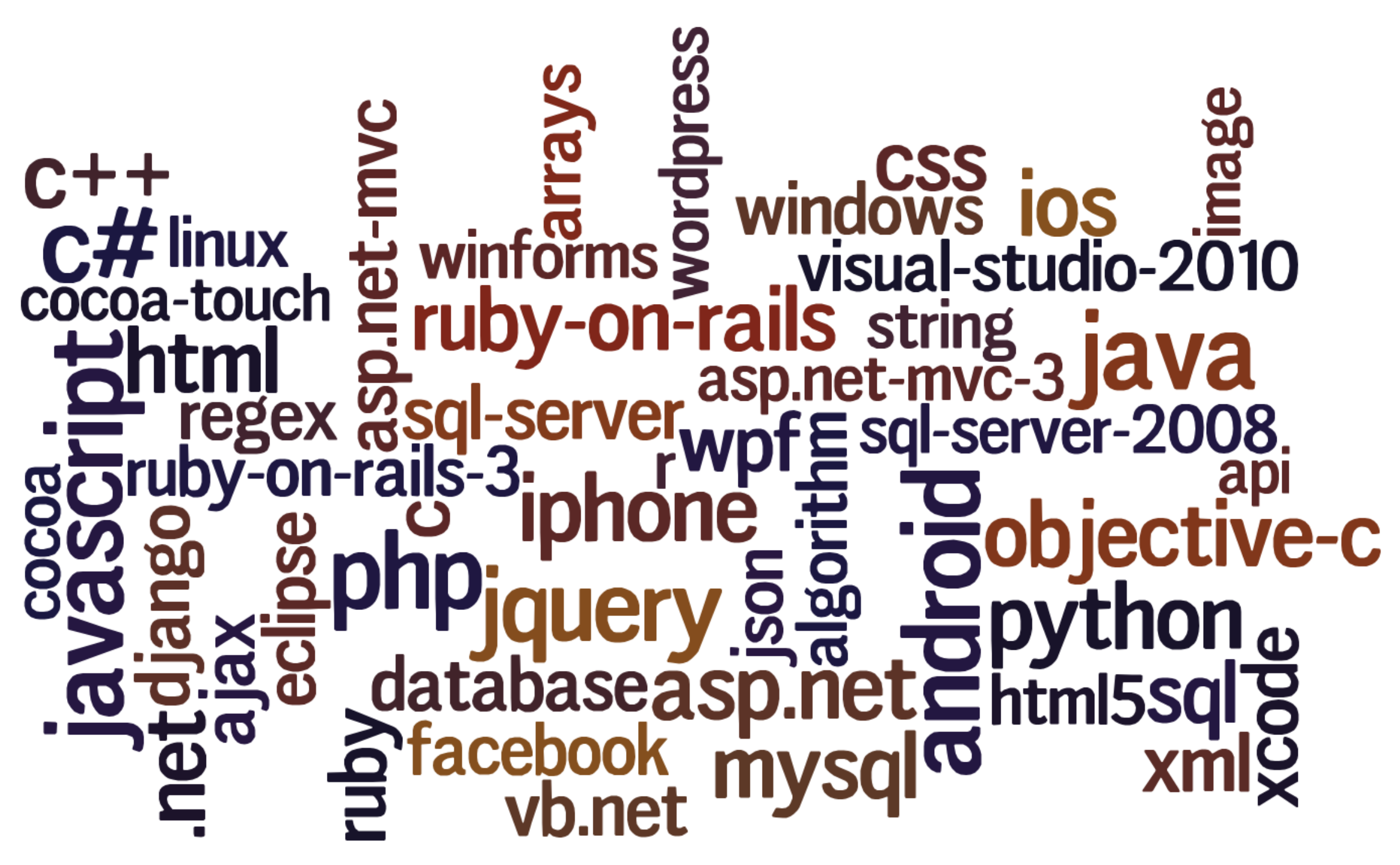}
\caption{shows the word cloud of the top-50 tags that occur in undeleted questions on Stack Overflow.}
\label{fig:tagcloud_undeleted_tags}
\end{figure}

We now analyze the \emph{tags} on \emph{undeleted} questions. Figure~\ref{fig:tagcloud_undeleted_tags} shows the word cloud of the top-50 tags that occur in undeleted questions on Stack Overflow. We notice the presence of programming related tags like \emph{objective-c}, \emph{android} and \emph{c\#} which points out these \emph{undeleted} questions are relevant to Stack Overflow.

\section{Deleted Question Prediction}

In the second phase of our study, we develop a predictive model to detect a deleted question at the time of question creation on Stack Overflow. We frame the problem of deleted question prediction as a binary classification task. In our supervised classification framework, we simulate real-world conditions viz. we only use information available at question creation time. We do not have access to information on answers and other crowdsourced information like \emph{view counts}, \emph{favorite} votes and \emph{question score}. In addition, Stack Overflow consists of millions of questions with thousands of topics (recall that there are 34,000+ \emph{tags}). We also perform our experiments on the entire unadulterated dataset of Stack Overflow. All these factors make the task of prediction of a deleted question at its creation time complex and extremely challenging in nature.

\subsection{Feature Identification}

We experiment with \textbf{47} features based on \emph{profile}($S_A$), \emph{community}($S_B$), \emph{question content}($S_C$) and \emph{text syntax}($S_D$) for our prediction task.~\emph{Profile} based features are based on the user-generated content on the Stack Overflow website. \emph{Community} based features are derived via the crowdsourced information generated by the Stack Overflow community. \emph{Question content} features are based on the text and metadata of the question while \emph{syntactic} or \emph{text syntax} features are based on the writing style of the text in the question. In order to investigate features based on question content, we make use of the latest Linguistic Inquiry and Word Count (LIWC2007) software~\cite{tausczik2010psychological}. LIWC2007 is a natural language psychometric analysis software that contains 4,500 words and 64 hierarchical dictionary categories. LIWC2007 takes a text document as input and outputs a score for the input over all 64 categories based on the writing style and psychometric properties of the document. We utilize the LIWC2007 software to understand the psychometric properties of natural language text in deleted questions. We find 11 LIWC categories -- personal pronouns(I, them, her), pronouns(I, them, itself), space(down, in, thin), relativity (area, bend, exit, stop), inclusive (and, with, include), cognitive process(cause, know, ought), social(mate, talk, they, child), function words, conjunctions (and, but, whereas) and prepositions(to, with, above) -- to be distinctive. We include these categories as features for our classification task. Table~\ref{tab:feat-set} shows all the 47 features based on four different categories. Features marked with $\dag$ are generated using LIWC2007.

\begin{table}[ht!] \scriptsize
\captionsetup{font=scriptsize, labelfont=bf, textfont=bf}
\centering
\caption{lists the 47 features used for our prediction task. Each feature belongs to a specific category and features marked with $\dag$ are generated using LIWC2007.}
\begin{tabular}{l|l|l} \hline
\textbf{Set} & \textbf{Type} & \textbf{Features}  \\ \hline
\multirow{12}{*}{$S_A$} & \multirow{12}{*}{Profile} & Age of Account \\
~ & ~ & Previous Questions with -ve score \\
~ & ~ & Previous Questions with +ve score \\
~ & ~ & Previous Questions with 0 score \\
~ & ~ & Previous Answers with -ve score \\
~ & ~ & Previous Answers with +ve score \\
~ & ~ & Previous Answers with 0 score \\
~ & ~ & Number of Previous Questions \\
~ & ~ & Number of Previous Answers  \\
~ & ~ & Number of Previous Badges    \\
~ & ~ & Questions to Age of Account Ratio  \\
~ & ~ & Answers to Age of Account Ratio  \\
\hline
\multirow{6}{*}{$S_B$} & \multirow{6}{*}{Community} & Average Answer Score \\
~ & ~ & Average Question View Counts \\
~ & ~ & Average Number of Comments \\
~ & ~ & Average Favorite Votes \\
~ & ~ & Average Question Score \\
~ & ~ & Average Number of Accepted Answers \\
\hline
\multirow{11}{*}{$S_C$} & \multirow{11}{*}{Content} & Number of URLs \\
~ & ~ & Number of Previous Tags \\
~ & ~ & Code Snippet Length \\
~ & ~ & LIWC score of Personal Pronouns$\dag$ \\
~ & ~ & LIWC score  of Pronouns$\dag$ \\
~ & ~ & LIWC score  of \emph{Space} words$\dag$ \\
~ & ~ & LIWC score  of \emph{Relativity} words$\dag$ \\
~ & ~ & LIWC score  of \emph{Inclusive} words$\dag$ \\
~ & ~ & LIWC score  of \emph{Cognitive Process} words$\dag$ \\
~ & ~ & LIWC score  of \emph{Social} words$\dag$ \\
~ & ~ & LIWC score  of 1st person singular pronouns$\dag$  \\
\hline
\multirow{18}{*}{$S_D$} & \multirow{18}{*}{Syntactic} & LIWC score of Function Words $\dag$\\
~ & ~ & LIWC score  of \emph{Conjunctions}$\dag$ \\
~ & ~ & LIWC score of Prepositions$\dag$ \\
~ & ~ & Number of characters in body \\
~ & ~ & Number of alphabetical characters in body \\
~ & ~ & Number of upper case characters in body \\
~ & ~ & Number of lower case characters in body \\
~ & ~ & Number of digit characters in body \\
~ & ~ & Number of white case characters in body \\
~ & ~ & Number of special characters in body \\
~ & ~ & Number of punctuation marks in body \\
~ & ~ & Number of words in body \\
~ & ~ & Number of short words in body \\
~ & ~ & Number of unique words in body \\
~ & ~ & Average body word length \\
~ & ~ & Number of characters in title \\
~ & ~ & Number of words in title \\
~ & ~ & Average title word length \\
\hline
\multicolumn{3}{c}{S$_A$=12, S$_B$=6, S$_C$=11, S$_D$=18, Total = 47 features} \\
\hline
\end{tabular}
\label{tab:feat-set}
\end{table}

Note that the challenge of our supervised learning task is to predict the probability of deletion at question creation time. Hence, we only consider features which are available at the time of question creation. We understand that other features, for example -- answer activity, could be a good discriminative feature for classification. But, including such features would violate the real-world conditions and therefore, we choose not to include such features in our experimental framework.

\subsection{Experimental Framework}

We recall that we have a total of 270,604 deleted questions as available by using the Stack Overflow database snapshots. 35,556 deleted questions do not have information about their question authors available. Since, an entire feature set in our supervised learning task is based on user profile we ignore these set of questions for this part of our study. Therefore, we have a final total of \textbf{235,048} deleted questions in the positive class of our dataset. We also see that the total number of non-deleted questions are extremely high in comparison to that of deleted questions viz. 95\% of the questions are non-deleted while only 5\% of these questions are deleted. Hence, the dataset is highly skewed towards the positive class. There have been various approaches in machine learning literature which deal with supervised learning for imbalanced class problems. One such popular approach is to randomly down sample the skewed or majority class data and make the dataset balanced~\cite{he2009learning}. Therefore, we randomly select \textbf{235,048} questions from the non-deleted class to form the negative class of our dataset. However, such a method may induce a sampling bias. In order to eliminate this bias, we draw 10 random samples of non-deleted questions and perform our prediction experiments on each random sample. We report our results on the average of the experiments drawn from all 10 random samples.

We experiment with multiple classifiers and find that \emph{Adaboost} classifier gives the best performance.~\emph{Adaboost} or Adaptive boosting is an ensemble based machine learning framework which combines the performance of a series of weak classifiers to configure a strong classifier~\cite{freund1995desicion}. Concretely, \emph{Adaboost} modifies subsequent classifiers in an attempt to correctly classify wrongly classified instances from previous classifiers. Prior work in content quality on community based question-answering websites have also observed best performance with ensemble-based learning methods~\cite{agichtein2008finding, Li:2012:APQ:2187980.2188200}. We use a 70-30\% training-testing split and perform 10-fold cross validation to prevent the problem of over fitting. We use decision tree as the base classifier and SAMME.R for the boosting algorithm~\cite{zhu2006multi}. The learning rate and number of estimators parameters are set to their default values. Table~\ref{tab:exp-setup} shows the experimental setup details for our supervised classification task.

\begin{table}[ht!] \scriptsize
\captionsetup{font=scriptsize, labelfont=bf, textfont=bf}
\centering
\caption{shows the experimental setup for the deleted question prediction task.}
\begin{tabular}{l|l} \hline
\textbf{Dataset} & 470,096 questions \\ 
\textbf{Deleted  (+ve class)} & 235,048 questions\\
\textbf{Non-Deleted (-ve class)} & 235,048 questions (10 times) \\
\textbf{Classifier} & Adaboost \\ 
\textbf{Learning Rate} & 1.0 \\ 
\textbf{Base Classifier} & Decision Tree \\
\textbf{Number of Estimators} & 100 \\
\textbf{Boosting Algorithm} & SAMME.R \\
\textbf{Cross Validation} & 10-fold \\
\multirow{2}{*}{\textbf{Classification Runs}} & 10-times \\
~ & (one for each +ve/-ve pair) \\
\textbf{Training-Testing split} & 70-30\% \\
\multirow{2}{*}{\textbf{Feature Sets}} & \{$S_A$\},\{$S_A$\, , $S_B$\}, \{$S_A$\,, $S_B$\,, $S_C$\},  \\
~ & \{$S_A$\,, $S_B$, $S_C$, $S_D$\} \\
\hline
\end{tabular}
\label{tab:exp-setup}
\end{table}

\subsection{Evaluation}

We now evaluate the performance of our classifier. Table~\ref{tab:conf-mat} shows the confusion matrix for our supervised classification task. We see that our system is able to accurately classify 65.9\% of deleted questions and 66.1\% of non-deleted questions with an overall accuracy of 66\%.

\begin{table}[ht!] \scriptsize
\captionsetup{font=scriptsize, labelfont=bf, textfont=bf}
\centering
\caption{Confusion Matrix -- Prediction Performance }
\begin{tabular}{cccc} \hline
~ & ~ & \multicolumn{2}{c}{\textbf{Predicted}} \\ \hline
~ & ~ & \textbf{Deleted} & \textbf{Non-Deleted} \\
\multirow{2}{*}{\textbf{True}} & \textbf{Deleted} & \textbf{65.9\%} & 34.1\% \\
~ & \textbf{Non-Deleted} & 33.9\% & \textbf{66.1\%} \\
\hline
\end{tabular}
\label{tab:conf-mat}
\end{table}

In order to understand the importance of our feature sets, we incrementally add feature sets and evaluate the performance of our classifier. Table~\ref{tab:eval-metric} shows the classification performance on incremental feature sets based upon multiple evaluation metrics -- F1 score, Accuracy and Area-Under-Curve(AUC). Notice the improvement in performance of our classifier on each feature set. This shows that our feature sets are important to the classification task.

\begin{table}[ht!] \scriptsize
\captionsetup{font=scriptsize, labelfont=bf, textfont=bf}
\centering
\caption{shows the classification performance on incremental feature sets based upon multiple evaluation metrics -- F1 score, Accuracy and Area-Under-Curve(AUC)}
\begin{tabular}{lccc} 
\textbf{Feature Set} & \textbf{F1} & \textbf{Accuracy} & \textbf{AUC} \\
\hline
\{$S_A$\} & 58.91 & 56.81 & 57.19 \\
\{$S_A$\, , $S_B$\} & 61.83 & 59.59 & 61.11 \\
\{$S_A$\,, $S_B$\,, $S_C$\} & 63.38 & 62.17 & 65.24 \\
\{$S_A$\,, $S_B$, $S_C$, $S_D$\} & 65.8 & 66.0 & 70.01 \\
\hline
\end{tabular}
\label{tab:eval-metric}
\end{table}

\subsection{Feature Analysis}

\begin{figure}[ht!]
\captionsetup{font=scriptsize, labelfont=bf, textfont=bf}
\centering
\includegraphics[width=\linewidth]{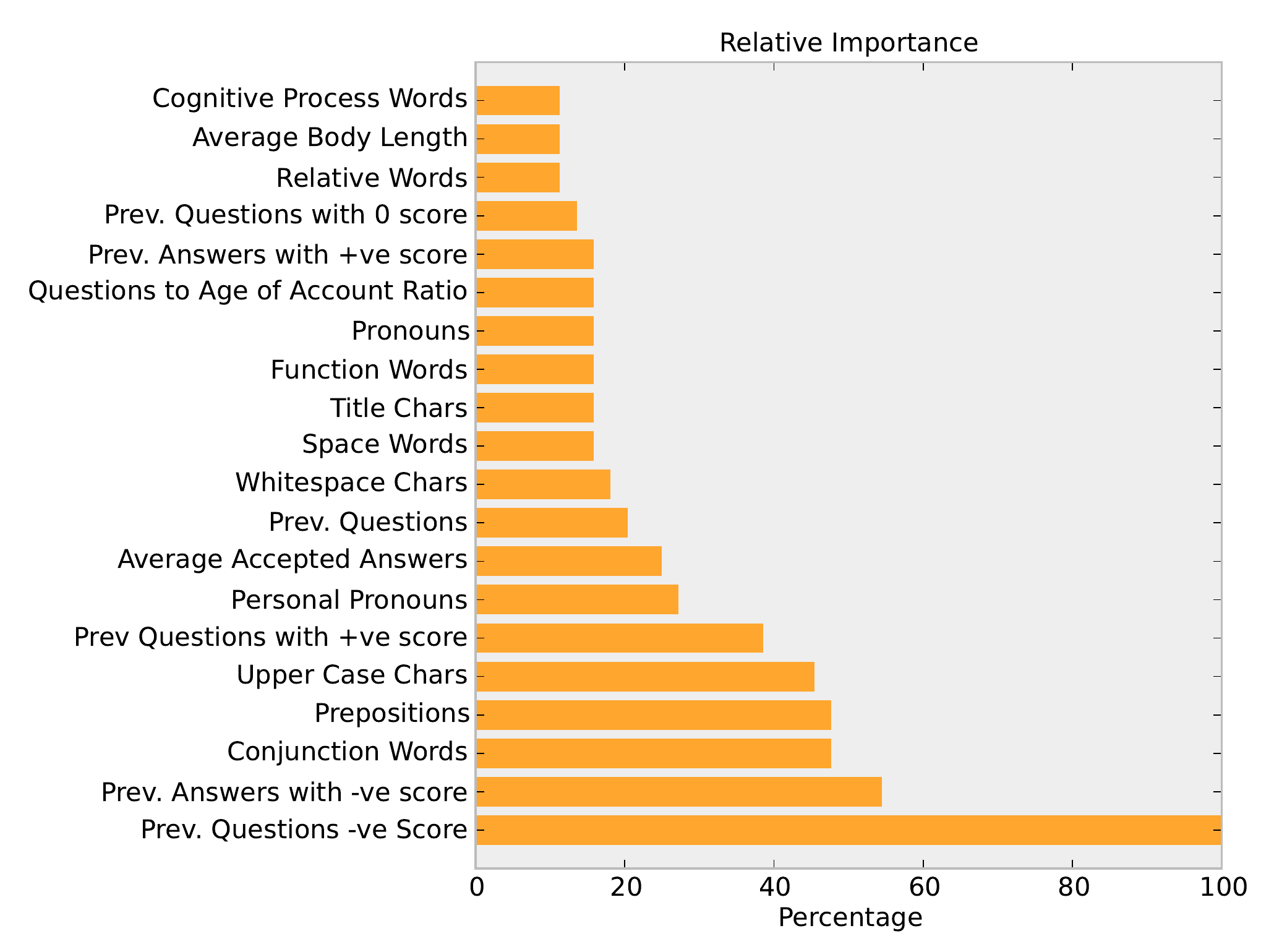}
\caption{shows the relative importance of the top 20 features for deleted question prediction.}
\label{fig:feat-imp}
\end{figure}

We now analyze our feature set in order to understand important features for deleted question prediction. The~\emph{Adaboost} classifier informs the discriminator power of each feature for classification. Figure~\ref{fig:feat-imp} shows the relative importance of top 20 features for deleted question prediction as given by our classifier. We see that the top-20 features are a mixture of features from all four categories of feature sets. It points out the importance of the use of different categories of feature sets for prediction.

\section{Conclusions}
We conduct the first large scale study of deleted questions on Stack Overflow.  We observe an increasing trend in the number of deleted questions on Stack Overflow over the last 2 years. The community takes significant time to detect a potential deleted question but moderators take swift appropriate action. However, we notice that moderators bear most of the encumbrance of detecting a deleted question. We also find that authors delete their own questions to salvage reputation points on the website. In general, deleted questions are extremely poor in worth to the Stack Overflow community. Also, deleted questions are significantly low in quality than `closed' questions. We discover the existence of a pyramidal structure of question quality in which deleted questions lie at the bottom of the pyramid (extremely poor quality). We also notice accidental question deletion by authors. We build a predictive model to detect deleted questions on Stack Overflow and report 66\% accurate predictions. We employ four categories of feature sets -- \emph{user profile}, \emph{community based}, \emph{content based} and \emph{stylistic} features -- and report most discriminatory features.


%
\bibliographystyle{abbrv}
\bibliography{delques}  

\begin{thebibliography}{10}

\bibitem{:2012zr}
Why and how are some questions deleted?
\newblock \url{http://stackoverflow.com/help/deleted-questions}.

\bibitem{:2008kx}
How does deleting work? what can cause a post to be deleted, and what does that
  actually mean? what are the criteria for deletion?
\newblock \url{http://meta.stackoverflow.com/q/5221/214223}, September 2008.

\bibitem{:2010kx}
The great question deletion audit of 2010.
\newblock
  \url{http://meta.stackoverflow.com/questions/51097/the-great-question-deletion-audit-of-2010},
  May 2010.

\bibitem{agichtein2008finding}
E.~Agichtein, C.~Castillo, D.~Donato, A.~Gionis, and G.~Mishne.
\newblock Finding high-quality content in social media.
\newblock In {\em Proceedings of the international conference on Web search and
  web data mining}, pages 183--194. ACM, 2008.

\bibitem{allamanis2013and}
M.~Allamanis and C.~Sutton.
\newblock Why, when, and what: analyzing stack overflow questions by topic,
  type, and code.
\newblock In {\em Proceedings of the Tenth International Workshop on Mining
  Software Repositories}, pages 53--56. IEEE Press, 2013.

\bibitem{andersonsteering}
A.~Anderson, D.~Huttenlocher, J.~Kleinberg, and J.~Leskovec.
\newblock Steering user behavior with badges.
\newblock 2013.

\bibitem{asaduzzaman2013answering}
M.~Asaduzzaman, A.~S. Mashiyat, C.~K. Roy, and K.~A. Schneider.
\newblock Answering questions about unanswered questions of stack overflow.
\newblock In {\em Proceedings of the Tenth International Workshop on Mining
  Software Repositories}, pages 97--100. IEEE Press, 2013.

\bibitem{Atwood:2009uq}
J.~Atwood.
\newblock Stack overflow creative commons data dump.
\newblock
  \url{http://blog.stackoverflow.com/2009/06/stack-overflow-creative-commons-data-dump/},
  June 2009.

\bibitem{barua2012developers}
A.~Barua, S.~W. Thomas, and A.~E. Hassan.
\newblock What are developers talking about? an analysis of topics and trends
  in stack overflow.
\newblock {\em Empirical Software Engineering}, pages 1--36, 2012.

\bibitem{bosu2013building}
A.~Bosu, C.~S. Corley, D.~Heaton, D.~Chatterji, J.~C. Carver, and N.~A. Kraft.
\newblock Building reputation in stackoverflow: an empirical investigation.
\newblock In {\em Proceedings of the Tenth International Workshop on Mining
  Software Repositories}, pages 89--92. IEEE Press, 2013.

\bibitem{Correa:2013vn}
D.~Correa and A.~Sureka.
\newblock Fit or unfit: Analysis and prediction of'closed questions' on stack
  overflow.
\newblock In {\em Proceedings of the ACM Conference on Online Social Networks},
  Boston, MA, USA, 2013. ACM.

\bibitem{freund1995desicion}
Y.~Freund and R.~E. Schapire.
\newblock A desicion-theoretic generalization of on-line learning and an
  application to boosting.
\newblock In {\em Computational learning theory}, pages 23--37. Springer, 1995.

\bibitem{gomez2013study}
C.~G{\'o}mez, B.~Cleary, and L.~Singer.
\newblock A study of innovation diffusion through link sharing on stack
  overflow.
\newblock In {\em Proceedings of the Tenth International Workshop on Mining
  Software Repositories}, pages 81--84. IEEE Press, 2013.

\bibitem{he2009learning}
H.~He and E.~A. Garcia.
\newblock Learning from imbalanced data.
\newblock {\em Knowledge and Data Engineering, IEEE Transactions on},
  21(9):1263--1284, 2009.

\bibitem{Jeff-Atwood:2009fk}
J.~S. Jeff~Atwood.
\newblock Stack exchange platform.
\newblock \url{http://stackexchange.com}, September 2009.

\bibitem{Jeon:2006:FPQ:1148170.1148212}
J.~Jeon, W.~B. Croft, J.~H. Lee, and S.~Park.
\newblock A framework to predict the quality of answers with non-textual
  features.
\newblock In {\em Proceedings of the 29th annual international ACM SIGIR
  conference on Research and development in information retrieval}, SIGIR '06,
  pages 228--235, New York, NY, USA, 2006. ACM.

\bibitem{Li:2012:APQ:2187980.2188200}
B.~Li, T.~Jin, M.~R. Lyu, I.~King, and B.~Mak.
\newblock Analyzing and predicting question quality in community question
  answering services.
\newblock In {\em Proceedings of the 21st international conference companion on
  World Wide Web}, WWW '12 Companion, pages 775--782, New York, NY, USA, 2012.
  ACM.

\bibitem{linares2013exploratory}
M.~Linares-V{\'a}squez, B.~Dit, and D.~Poshyvanyk.
\newblock An exploratory analysis of mobile development issues using stack
  overflow.
\newblock In {\em Proceedings of the Tenth International Workshop on Mining
  Software Repositories}, pages 93--96. IEEE Press, 2013.

\bibitem{mamykina2011design}
L.~Mamykina, B.~Manoim, M.~Mittal, G.~Hripcsak, and B.~Hartmann.
\newblock Design lessons from the fastest q\&a site in the west.
\newblock In {\em Proceedings of the 2011 annual conference on Human factors in
  computing systems}, pages 2857--2866. ACM, 2011.

\bibitem{nasehi2012makes}
S.~M. Nasehi, J.~Sillito, F.~Maurer, and C.~Burns.
\newblock What makes a good code example?: A study of programming q\&a in
  stackoverflow.
\newblock In {\em Software Maintenance (ICSM), 2012 28th IEEE International
  Conference on}, pages 25--34. IEEE, 2012.

\bibitem{pal2012exploring}
A.~Pal, F.~M. Harper, and J.~A. Konstan.
\newblock Exploring question selection bias to identify experts and potential
  experts in community question answering.
\newblock {\em ACM Transactions on Information Systems (TOIS)}, 30(2):10, 2012.

\bibitem{ponzanelli2013seahawk}
L.~Ponzanelli, A.~Bacchelli, and M.~Lanza.
\newblock Seahawk: stack overflow in the ide.
\newblock In {\em Proceedings of the 2013 International Conference on Software
  Engineering}, pages 1295--1298. IEEE Press, 2013.

\bibitem{shah2010evaluating}
C.~Shah and J.~Pomerantz.
\newblock Evaluating and predicting answer quality in community qa.
\newblock In {\em Proceedings of the 33rd international ACM SIGIR conference on
  Research and development in information retrieval}, pages 411--418. ACM,
  2010.

\bibitem{tausczik2010psychological}
Y.~R. Tausczik and J.~W. Pennebaker.
\newblock The psychological meaning of words: Liwc and computerized text
  analysis methods.
\newblock {\em Journal of Language and Social Psychology}, 29(1):24--54, 2010.

\bibitem{wang2013detecting}
W.~Wang and M.~W. Godfrey.
\newblock Detecting api usage obstacles: a study of ios and android developer
  questions.
\newblock In {\em Proceedings of the Tenth International Workshop on Mining
  Software Repositories}, pages 61--64. IEEE Press, 2013.

\bibitem{zhu2006multi}
J.~Zhu, S.~Rosset, H.~Zou, and T.~Hastie.
\newblock Multi-class adaboost.
\newblock {\em Ann Arbor}, 1001(48109):1612, 2006.

\end{thebibliography}
%
%

\end{document}